\pdfoutput=1

\documentclass[aps,prc,reprint,groupedaddress,nofootinbib]{revtex4-1}

\def\D{\mathrm{d}}
 
\newcommand{\pT}{\ensuremath{p_\mathrm{T}}}
\newcommand{\mT}{\ensuremath{m_\mathrm{T}}}

\usepackage{tikz}
\usepackage{graphicx}
\usepackage{units}
\usepackage{amsmath}

\usepackage[colorlinks=true,linkcolor=blue,citecolor=blue, urlcolor=blue]{hyperref}   

    \newcommand{\kompost}{K{\o}MP{\o}ST}

\def\Eq#1{Eq.~(\ref{#1})}
\def\Eqs#1{Eqs.~(\ref{#1})}
\def\eq#1{(\ref{#1})}
\def\app#1{Appendix~\ref{#1}}
\def\Fig#1{Fig.~\ref{#1}}
\def\Figs#1{Figs.~\ref{#1}}
\def\Sec#1{Sec.~\ref{#1}}
\def\App#1{Appendix~\ref{#1}}
\newcommand{\subfig}[2]{%
\begin{tikzpicture}%
\node[rectangle] (image) at (0,0) {#2};
\node[anchor=south west] (label) at (image.south west) {(#1)};
\end{tikzpicture}%
}
\begin{document}

\title{Entropy production in pp and Pb-Pb collisions at energies available at
    the CERN Large Hadron Collider}

\author{Patrick Hanus}
\affiliation{Physikalisches Institut, Universit\"at Heidelberg, Im Neuenheimer Feld 226, D-69120 Heidelberg, Germany}
\author{Aleksas Mazeliauskas}
\affiliation{Theoretical Physics Department, CERN, CH-1211 Gen\`eve 23, Switzerland}
\affiliation{Institut f\"ur Theoretische Physik, Universit\"at Heidelberg, Philosophenweg 16, D-69120 Heidelberg, Germany}
\author{Klaus Reygers}
\affiliation{Physikalisches Institut, Universit\"at Heidelberg, Im Neuenheimer Feld 226, D-69120 Heidelberg, Germany}

\date{\today}

\newcommand\dSdySP{11\,335}
\newcommand\dSdySPErr{1188}
\newcommand\dSdySPErrSpectra{629}
\newcommand\dSdySPErrHBTRadii{1007}
\newcommand\SoverNchSP{6.7}
\newcommand\SoverNchSPErr{0.8}

\newcommand\dSdySPPPMinBias{37.8}
\newcommand\dSdySPPPMinBiasErr{3.7}
\newcommand\dSdySPPPMinBiasErrSpectra{3.3}
\newcommand\dSdySPPPMinBiasErrHBTRadii{1.8}
\newcommand\SoverNchSPPPMinBias{5.2}
\newcommand\SoverNchSPErrPPMinBias{0.5}

\newcommand\dSdySPPPHighMult{135.7}
\newcommand\dSdySPPPHighMultErr{17.9}
\newcommand\dSdySPPPHighMultErrSpectra{12.3}
\newcommand\dSdySPPPHighMultErrHBTRadii{13.1}
\newcommand\SoverNchSPPPHighMult{5.4}
\newcommand\SoverNchSPErrPPHighMult{0.7}

\newcommand\NoverNch{1.115}
\newcommand\NoverNchErr{0.03}

\newcommand\dSdyGen{11\,534}
\newcommand\dSdyGenErr{1188}
\newcommand\SoverNchGen{6.9}
\newcommand\SoverNchGenErr{0.8}

\newcommand\dSdyGenPPMinBias{41.7}
\newcommand\dSdyGenPPMinBiasErr{4.1}
\newcommand\SoverNchGenPPMinBias{5.7}
\newcommand\SoverNchGenErrPPMinBias{0.6}

\newcommand\dSdyGenPPHighMult{159.0}
\newcommand\dSdyGenPPHighMultErr{19.8}
\newcommand\SoverNchGenPPHighMult{6.3}
\newcommand\SoverNchGenErrPPHighMult{0.8}

\def\Eq#1{Eq.~(\ref{#1})}
\def\Eqs#1{Eqs.~(\ref{#1})}
\def\eq#1{(\ref{#1})}
\def\app#1{Appendix~\ref{#1}}
\def\Fig#1{Fig.~\ref{#1}}
\def\Figs#1{Figs.~\ref{#1}}
\def\Sec#1{Sec.~\ref{#1}}
\def\App#1{Appendix~\ref{#1}}

\begin{abstract}
We use experimentally measured identified particle spectra and Hanbury
Brown-Twiss radii to determine the entropy per unit rapidity $\D S/\D y$
produced in $\sqrt{s} = \unit[7]{TeV}$ pp and $\sqrt{s_\mathrm{NN}} =
\unit[2.76]{TeV}$ Pb--Pb collisions. We find that $\D S/\D y = \dSdySP \pm
\dSdySPErr$ in 0--10\% Pb--Pb, $\D S/\D y = \dSdySPPPHighMult \pm
\dSdySPPPHighMultErr$ in high-multiplicity pp, and $\D S/\D y =
\dSdySPPPMinBias \pm \dSdySPPPMinBiasErr$ in minimum bias pp collisions and
compare the corresponding entropy per charged particle $(\D S/\D y)/(\D N_\mathrm{ch} /\D y)$ to
predictions of statistical models. Finally, we use the QCD kinetic
theory pre-equilibrium and viscous hydrodynamics to model entropy production in the collision and reconstruct
the average temperature profile at $\tau_0\approx \unit[1]{fm}/c$ for high multiplicity pp and Pb--Pb collisions.

\end{abstract}

\pacs{}

\maketitle

\section{Introduction} 

Ultrarelativistic collisions of nuclei as studied at RHIC and the LHC are
typically modeled assuming rapid thermalization within a time scale of
\unit[1--2]{fm}/$c$~\cite{Busza:2018rrf}. The subsequent longitudinal and
transverse expansion of the created quark-gluon plasma (QGP) is then described
by viscous relativistic hydrodynamics~\cite{deSouza:2015ena}. In this picture
the bulk of the entropy is created during the thermalization process and the
later stages of the evolution add relatively little~\cite{Muller:2011ra}. By
correctly accounting for the entropy production in different stages of the
collisions, one can therefore relate the measurable final-state particle
multiplicities to the properties of system, e.g.\ initial temperature, at the
earlier stages of the collisions.

Two different methods are frequently used to estimate the total produced
entropy in nuclear collisions. In the first method, pioneered by Pal and
Pratt, one calculates the entropy based on transverse momentum spectra of
different particle species and their source radii as determined from Hanbury
Brown-Twiss correlations~\cite{Pal:2003rz}. The original paper analyzed data
from $\sqrt{s_\mathrm{NN}} = \unit[130]{GeV}$  Au--Au collisions and is still
the basis of many entropy estimations at other energies~\cite{Muller:2011ra}.
The second method uses the entropy per hadron as calculated in a hadron
resonance gas model to translate the  final-state multiplicity $\D N/\D y$ per
unit of rapidity to an entropy
d$S$/d$y$~\cite{Sollfrank:1992ru,Muller:2005en}. Even though the estimate of
the entropy from the measured multiplicity $\D N_\mathrm{ch}/\D\eta$ is
relatively straightforward one finds quite different values for the conversion
factor between the measured charged-particle multiplicity $\D
N_\mathrm{ch}/\D\eta$ and the entropy $\D S/\D y$ in the
literature~\cite{Gubser:2008pc,Nonaka:2005vr,Muller:2005en,Berges:2017eom}.

This paper provides an up-to-date calculation of entropy production in pp and
Pb--Pb collisions at the LHC energies and uses state-of-the-art modeling of the QGP
to reconstruct the initial conditions at the earliest moments in the
collision. In \Sec{sec:method} we recap the method of Ref.~\cite{Pal:2003rz},
which we use in \Sec{sec:PbPb} and \Sec{sec:pp} to calculate the total
produced entropy per rapidity, and the entropy per final-state charged hadron
$S/N_\mathrm{ch} \equiv (\D S/\D
y) / (\D   N_\mathrm{ch}/\D y)$ from the identified particle spectra and
femtoscopy data for  $\sqrt{s} = \unit[7]{TeV}$ pp and
$\sqrt{s_\mathrm{NN}} =\unit[2.76]{TeV}$ Pb--Pb collisions at
LHC~\cite{Abelev:2013vea,Abelev:2013xaa,ABELEV:2013zaa,Adam:2015qaa,Abelev:2013xaa,Abelev:2012jp,Acharya:2018orn,ALICE:2017jyt}. 
In \Sec{sec:hrg} the result for the entropy per particle is then compared to
different estimates of the entropy per hadron calculated in hadron resonance
gas models at the chemical freeze-out
temperature of $T_\mathrm{ch} \approx \unit[156]{MeV}$~\cite{Andronic:2017pug}.
Finally in \Sec{sec:IC} we use different models of the QGP evolution to track
entropy production in different stages of the collisions and to determine the
initial temperature profile at $\tau= \unit[1]{fm}/c$.

\section{Entropy from transverse momentum spectra and HBT radii\label{sec:method}} 
In this section we recap the entropy calculation from phase-space densities
obtained from particle spectra and femtoscopy~\cite{Pal:2003rz}. Foundations
for this method were laid in  Refs.~\cite{Bertsch:1994qc, Ferenc:1999ku}. The entropy $S$
for a given hadron species at the time of kinetic freeze-out is calculated
from the phase space density $f(\vec p, \vec r)$ according to
\begin{align} 
  S &= (2J + 1) 
  \int \frac{\D^3 r \D^3 p}{\left(2 \pi \right)^3} [ -f \ln
  f \pm \left(1 \pm f \right) \ln \left(1 \pm f \right) ]
\label{eq:entropy}
\end{align}
where $+$ is for bosons and the $-$ for fermions. The factor $2J+1$ is the spin degeneracy. 
The total entropy in the collision is then given by the sum of the entropies of
the produced hadrons species. The integral in \Eq{eq:entropy} can be evaluated 
using the series expansion
\begin{align}
	\pm \left(1 \pm f \right) \ln \left(1 \pm f \right) =  
	f \pm \frac{f^2}{2} - \frac{f^3}{6} \pm \frac{f^4}{12} + \ldots\,.
	\label{eq:taylor}
\end{align} 
Three-dimensional source radii measured through Hanbury Brown-Twiss
two-particle correlations \cite{Lisa:2005dd} are usually determined in
the longitudinally co-moving system (LCMS) in which the component of 
the pair momentum along the beam direction vanishes.  
The density profile of the source in the LCMS is parametrized by a three-dimensional 
Gaussian so that the phase space density can be written as 
\begin{equation}
f(\vec{p},\vec{r}) = \mathcal{F}(\vec{p}) \exp{\left(
-\frac{x_\mathrm{out}^2}{2 R^2_\mathrm{out}} 
-\frac{x_\mathrm{side}^2}{2 R^2_\mathrm{side}}
-\frac{x_\mathrm{long}^2}{2 R^2_\mathrm{long}}
\right)} \label{eq:f_Pratt} 
\end{equation} 
with
\begin{equation} \mathcal{F}(\vec{p}) = \frac{(2 \pi)^{3/2}}{2J+1}
\frac{\D^3 N}{\D^3 p} \frac{1}{R_\mathrm{out} R_\mathrm{side} R_\mathrm{long}}.  
\label{eq:fmax}
\end{equation} 
The radii in Eqs.~(\ref{eq:f_Pratt}) and (\ref{eq:fmax}) are functions of the 
momentum $\vec{p}$. 

In many cases only the one-dimensional source radius $R_\mathrm{inv}$, which
is determined in the pair rest frame (PRF), can be determined experimentally
owing to limited statistics. In Ref.~\cite{Pal:2003rz} the relation between
$R_\mathrm{inv}$ in the PRF and the three-dimensional radii in the LCMS was
assumed to be
\begin{equation}
R^3_\mathrm{inv} \approx
\gamma R_\mathrm{out} R_\mathrm{side} R_\mathrm{long},
\label{eq:Rinv} 
\end{equation}
where $\gamma = m_\mathrm{T}/m \equiv \sqrt{m^2 + p_\mathrm{T}^2}/m$. 
This is also our standard assumption. In \cite{Adam:2015vna} and
\cite{Aamodt:2011kd} the ALICE collaboration reported values for both
$R_\mathrm{inv}$ and $R_\mathrm{out}$, $R_\mathrm{side}$, $R_\mathrm{long}$
obtained from two-pion correlations in Pb--Pb collisions at
$\sqrt{s_\mathrm{NN}} = \unit[2.76]{TeV}$ and pp collisions at $\sqrt{s} =
\unit[7]{TeV}$, respectively. From these results one can determine a more
general version of \Eq{eq:Rinv} of the form $R^3_\mathrm{inv} \approx
h(\gamma) R_\mathrm{out} R_\mathrm{side} R_\mathrm{long}$ with $h(\gamma) =
\alpha \gamma^\beta$. Results for the entropy $\D S/\D y$ obtained under this
assumption are given in \app{sec:RinvRoutRsideRlong}.

Using \Eq{eq:Rinv} one arrives at
\begin{multline} 
\frac{\D S}{\D y} = \int
\D p_T \ 2 \pi  p_T \  E\frac{\D^3 N}{\D^3 p}  \left(\frac{5}{2} - 
\ln \mathcal{F} \pm \frac{\mathcal{F}}{2^{5/2}} \right. \\ 
- \left. \frac{\mathcal{F}^2}{2 \cdot 3^{5/2}} \pm \frac{\mathcal{F}^3}{3 \cdot 4^{5/2}} 
\right)
\label{eq:dSdy}
\end{multline} 
with 
\begin{equation} 
\mathcal{F} = \frac{1}{m} \frac{(2 \pi)^{3/2}}{2J+1} 
\frac{1}{R^3_\mathrm{inv}(m_\mathrm{T})} E\frac{\D^3 N}{\D^3 p} 
\end{equation} 
where $m$ is the particle mass and $+$ is for bosons and $-$ for fermions. Note that
\Eq{eq:dSdy} includes the terms up to $f_i^4/12$ of the Taylor expansion in
\Eq{eq:taylor}. 

Pions have the highest phase space density of the considered hadrons and the
approximation made in \Eq{eq:dSdy} is better than 1\% for pions in
central Pb--Pb collisions at $\sqrt{s_\mathrm{NN}} = \unit[2.76]{TeV}$. In pp
collisions at $\sqrt{s} = \unit[7]{TeV}$ the maximum pion phase space density
$\mathcal{F}(p_\mathrm{T})$ exceeds unity at low $p_\mathrm{T}$ rendering the
series expansion in \Eq{eq:taylor} unreliable. For pions in pp
collisions we therefore approximate the $(1+f) \ln (1+f)$ term of
\Eq{eq:entropy} by a polynomial of order 8. This gives an approximate
expression with numerical coefficients $a_i$ which is also valid for values of
$\mathcal{F}$ obtained for pions in high-multiplicity pp collisions:
\begin{equation} 
\frac{\D S}{\D y} = \int
\D p_T \ 2 \pi  p_T \  E\frac{\D^3 N}{\D^3 p}  \left(\frac{5}{2} - 
\ln \mathcal{F} + \sum_{i=0}^7 a_i \mathcal{F}^i \right).
\label{eq:dSdyPoly}
\end{equation}

\section{Results}
\subsection{Entropy in Pb--Pb collisions at \boldmath{$\sqrt{s_\mathrm{NN}} = 2.76\,\mathrm{TeV}$}\label{sec:PbPb}} 
We determine the entropy in Pb--Pb collisions at $\sqrt{s_\mathrm{NN}} =
\unit[2.76]{TeV}$ for the 10\% most central collisions considering as
final-state hadrons the particles given in Table~\ref{tab:dsdy}. The
calculation uses transverse momentum spectra of $\pi$, $\mathrm{K}$,
$\mathrm{p}$ \cite{Abelev:2013vea}, $\Lambda$ \cite{Abelev:2013xaa}, and
$\Xi$, $\Omega$ \cite{ABELEV:2013zaa} from the ALICE collaboration as
experimental input. We also use HBT radii measured by ALICE
\cite{Adam:2015vja}.

For the entropy determination the measured transverse momentum spectra need to
be extrapolated to $\pT = 0$. To this end we fit different functional forms to the
$\pT$ spectra (Tsallis, Bose-Einstein, exponential in transverse mass $\mT =
\sqrt{\pT^2 + m^2}$, Boltzmann, as defined in \cite{Abelev:2013vea}). In the
entropy calculation we only use the extrapolations in $\pT$ regions where data
are not available, otherwise we used the measured spectra. Differences of the
entropy estimate for different functional form are taken as a contribution to
the systematic uncertainty. We have checked that the $\pT$-integrated $\pi$,
$\mathrm{K}$, $\mathrm{p}$ multiplicities $(\D n/\D y)_{y=0}$ agree with the
values published in \cite{Abelev:2013vea}.

\begin{figure} 
\includegraphics[width=0.95\linewidth]{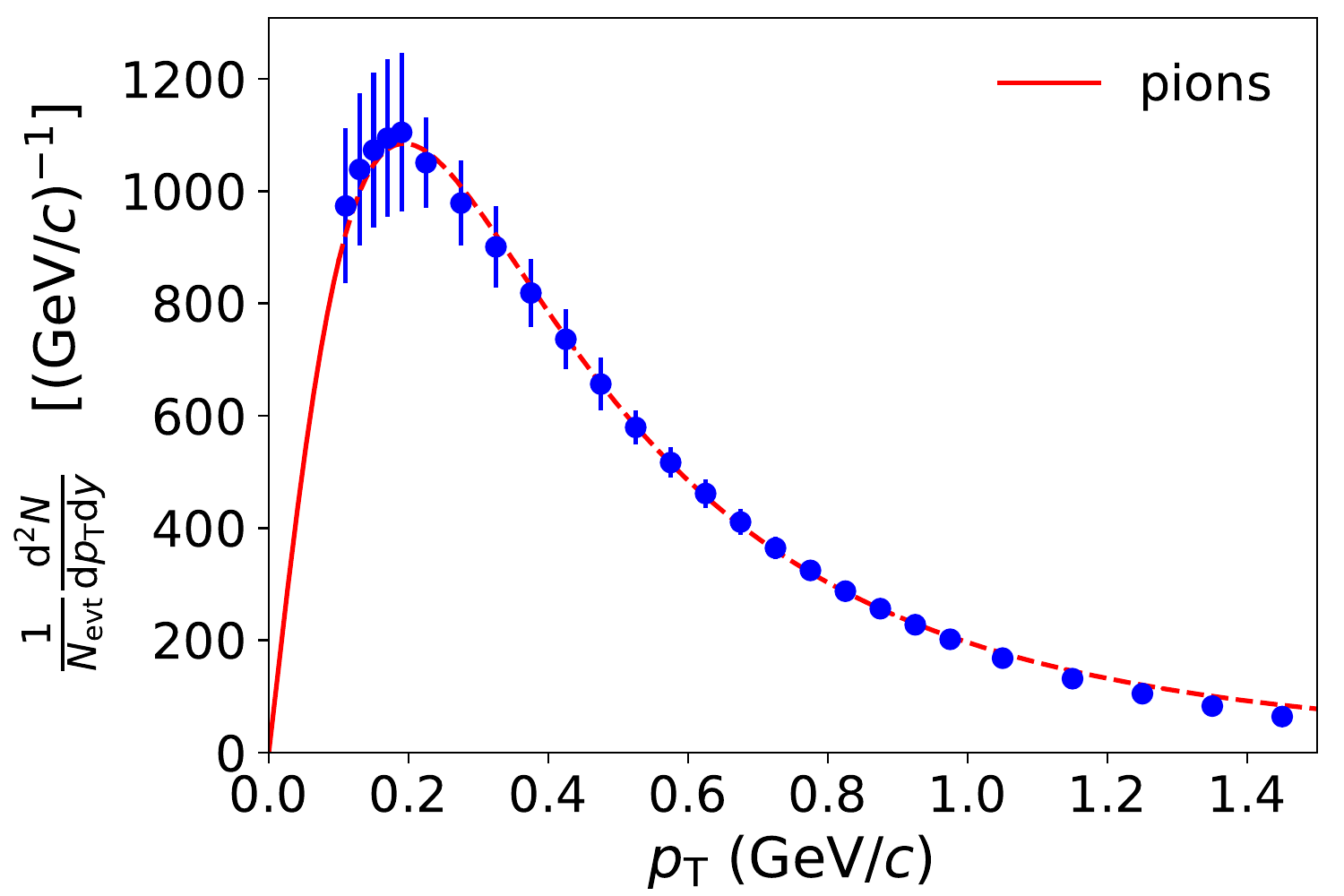}
\includegraphics[width=0.95\linewidth]{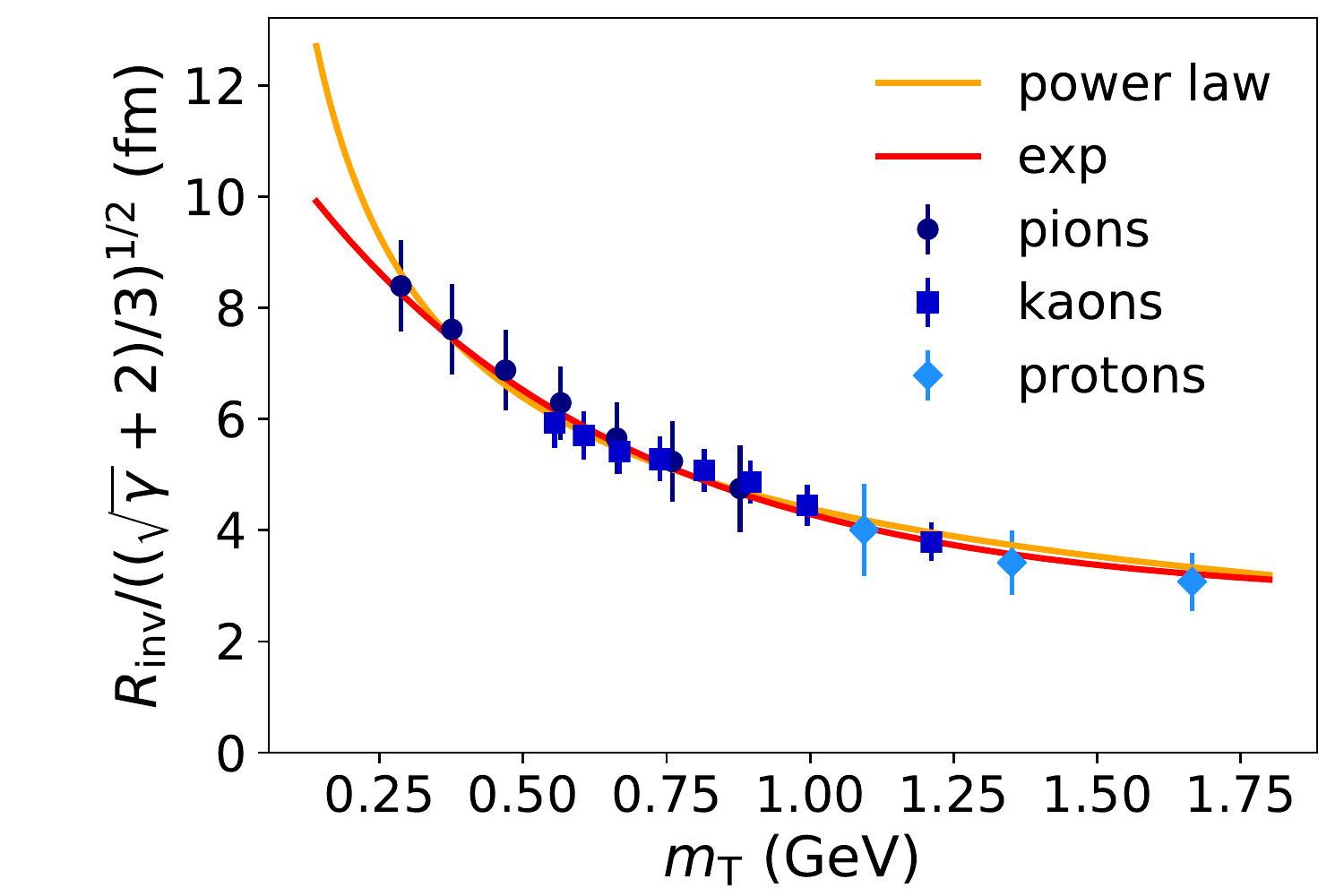}
\caption{\label{fig:pion_spectrum_and_Rinv}Transverse momentum spectrum of positive 
pions (top) and scaled HBT radii $R_\mathrm{inv}$ (bottom) in 0--10\%  Pb--Pb collisions at
$\sqrt{s_\mathrm{NN}} = \unit[2.76]{TeV}$. A Tsallis function
\cite{Tsallis:1987eu,Bhattacharyya:2017hdc} is fitted to the spectrum to
extrapolate to $\pT = 0$. The one-dimensional HBT radii divided by
$((\sqrt{\gamma}+2)/3)^{1/2}$ \cite{Kisiel:2014upa} where
$\gamma = m_\mathrm{T}/m$ as a function of the transverse mass
$m_\mathrm{T}$ are parametrized by a power law function $\alpha
m_\mathrm{T}^\beta$ and by an exponential function $a \exp(-m_\mathrm{T}/b) + c$.} 
\end{figure} 
 
The one-dimensional invariant HBT radii $R_\mathrm{inv}$ are only available
for $\pi$, $\mathrm{K}$, and $\mathrm{p}$. When plotted as a function of
transverse mass $m_\mathrm{T} = \sqrt{m^2 + p_\mathrm{T}^2}$ the
$R_\mathrm{inv}$ values for these particles do not fall on a common curve.
However, in \cite{Kisiel:2014upa} it was shown that the HBT radii
$R_\mathrm{inv}$ divided by $((\sqrt{\gamma}+2)/3)^{1/2}$ where $\gamma =
m_\mathrm{T}/m$ are approximately a function of $m_\mathrm{T}$ only. This
empirical scaling factor for $R_\mathrm{inv}$ is related to the fact that for
a three-dimensional Gaussian parametrization of the source the
one-dimensional source distribution in general cannot be described by a Gaussian
(see Appendix of \cite{Kisiel:2014upa}). We use this $m_\mathrm{T}$ scaling of the
scaled HBT radii to obtain $R_\mathrm{inv}(m_\mathrm{T})$ for all considered
particles. The bottom panel of Fig.~\ref{fig:pion_spectrum_and_Rinv} shows
parametrizations of the scaled HBT radii with a power law function and with an
exponential function which provide different extrapolation towards the pion
mass.
We propagate the systematic uncertainties of the measured HBT radii as well
as the uncertainty related to the two different parametrizations to the
uncertainty of the extracted entropy. 

For the entropy calculation the particle species in Table~\ref{tab:dsdy} are
considered stable. The entropy carried by neutrons, neutral kaons, $\eta$,
$\eta'$, and $\Sigma$ baryons is estimated based on measured species
assuming that the entropy per particle is similar for particles with similar
masses. The entropy carried by neutrons is assumed to be the same as the
entropy carried by protons. The entropy associated with neutral kaons and
$\eta$ mesons is determined from charged kaons, the entropy of $\eta'$ from
protons, and the entropy of $\Sigma$ baryons from $\Lambda$. 

The yields of particles in Table~\ref{tab:dsdy} contain contributions from
strong decays. To take into account mass differences and to estimate the
contributions from strong decays to the different particle species we simulate
particle decays with the aid of Pythia 8.2
\cite{Sjostrand:2006za,Sjostrand:2007gs}. To this end we generate primary
particles available in Pythia 8.2 with rates proportional to equilibrium
particle densities in a noninteracting hadron gas
\cite{BraunMunzinger:2003zd,Letessier:2002gp}:
\begin{equation} 
n = \sum_{k=1}^{\infty} \frac{T g}{2 \pi^2} \frac{(\pm)^{k+1}}{k} m^2 \mathrm{K}_2
\left(\frac{k m}{T}\right) e^{k \mu / T}.  
\label{eq:particle_density}
\end{equation} 
Here $g = 2J+1$ is the spin degeneracy factor and $K_2$ the modified Bessel
function of the second kind. The $+$ is for bosons and the $-$ for fermions.
For the chemical potential we use $\mu = 0$. For the temperature we take $T =
\unit[156]{MeV}$ as obtained from statistical model fits to particle yields
measured at the LHC \cite{Andronic:2017pug}. We then simulate strong and
electromagnetic decays of the primary particles. Particle ratios after decays
are used to estimate the entropy of unmeasured particles. In case of the
$\eta$ meson we find that after decays the $\eta/\mathrm{K}^+$ ratio is 0.69
while the primary ratio is $\eta_\mathrm{prim}/\mathrm{K}^+_\mathrm{prim} =
0.79$. For the $\eta'$ we find $\eta'_\mathrm{prim}/\mathrm{p}_\mathrm{prim} =
0.45$ and $\eta'/\mathrm{p} = 0.25$ after decays. The primary
$\Sigma_\mathrm{prim}^-/\Lambda_\mathrm{prim}$ ratio is about $0.66$. The
entropy carried by the $\Sigma$ baryons is derived from the ratios
$\Sigma^-/\Lambda \approx 0.26$ and $\Sigma^0/\Lambda \approx 0.27$ after
decays.

The $\eta$, $\eta'$ mesons and $\Sigma^0$ baryons decay electromagnetically.
Decay products from these decays ($\eta, \eta' \to \mathrm{pions}$, and
$\Sigma^0 \to \Lambda \gamma$) are not subtracted from the experimentally
determined particle spectra. As $\eta$, $\eta'$, and $\Sigma$ are considered
stable in the entropy calculation (see Table~\ref{tab:dsdy}) we correct for
this feeddown contributions. In the particle decay simulation described above
we determine the feeddown fraction 
\begin{equation} 
R_\mathrm{fd}(X \to Y) =
\frac{\mbox{number of}\;Y\; \mbox{from}\;X}{\mbox{total number of}\;Y}
\end{equation} 
and find $R_\mathrm{fd}(\eta \to \pi^+) = 3.6\%$,
$R_\mathrm{fd}(\eta' \to \pi^+) = 1.2\%$, $R_\mathrm{fd}(\eta' \to \eta) = 5.9\%$, 
and $R_\mathrm{fd}(\Sigma^0 \to
\Lambda) = 27.0\%$. 

The entropies for the particle species considered stable are summarized in
Table~\ref{tab:dsdy}. These values represent the average of the entropies
obtained for the power law and the exponential parametrization of the scaled
invariant HBT radii. In both cases the  Tsallis function was used to
extrapolate the measured transverse momentum spectra to $p_\mathrm{T} = 0.$
We considered the uncertainties of the measured transverse momentum spectra,
the choice of the parametrization of the $p_\mathrm{T}$ spectra, the
uncertainties of the measured HBT radii, and the choice of the parametrization
of the HBT radii as a function of $m_\mathrm{T}$ as sources of systematic
uncertainties. The estimated total entropy in 0--10\% most central Pb--Pb
collisions at $\sqrt{s_\mathrm{NN}} = \unit[2.76]{TeV}$ is $\dSdySP \pm
\dSdySPErr$. The uncertainty of the estimated entropy is the quadratic sum of
the uncertainties related to the transverse momentum spectra
($\sigma_\mathrm{spectra} = \dSdySPErrSpectra$) and invariant HBT radii
($\sigma_{R_\mathrm{inv}} = \dSdySPErrHBTRadii$).  
\begin{table}[h]
\centering
\caption{Estimate of the entropy $(\D S/\D y)_{y=0}$ for 0--10\% most central
Pb--Pb collisions at $\sqrt{s_\mathrm{NN}} = \unit[2.76]{TeV}$. The table
shows the hadrons considered as stable final-state particles and their
contribution to the total entropy.\label{tab:dsdy}} 
\begin{tabular}{@{}c @{\hskip 1em} c @{\hskip 1em} c @{\hskip 1em} c@{}} 
particle &  $(\D S/\D y)_{y=0}^{\text{one state}}$ & factor &  $(\D S/\D y)_{y=0}^{\text{total}}$\\
\hline \hline 
$\pi$ & 2182 & 3 & 6546 \\
$\mathrm{K}$ & 605 & 4 & 2420 \\ 
$\eta$ & 399 & 1 & 399 \\ 
$\eta'$ &  66& 1 &  66 \\ 
$\mathrm{p}$ & 266 & 2 & 532 \\ 
$\mathrm{n}$ & 266 & 2 & 532 \\
$\Lambda$ & 160 & 2 & 320 \\ 
$\Sigma$ & 58 & 6 & 348 \\ 
$\Xi$ & 39 & 4 & 156 \\ 
$\Omega$ & 8 & 2 & 16 \\ 
\hline 
total &  &  & \dSdySP \\
\end{tabular}
\end{table}
 
It is interesting to calculate the entropy per charged hadron in the final
state from the total entropy. From \cite{Aamodt:2010cz} we obtain for 0--10\%
most central Pb--Pb collisions at $\sqrt{s_\mathrm{NN}} = \unit[2.76]{TeV}$ a
charged-particle multiplicity at mid-rapidity of $\D N_\mathrm{ch}/\D\eta =
1448 \pm 54$. From our parametrizations of the pion, kaon, and proton spectra
we find a Jacobian for the change of variables from pseudorapidity to rapidity
of $(\D N_\mathrm{ch}/\D y) / (\D N_\mathrm{ch}/\D\eta) = 1.162 \pm 0.008$.
This yields an entropy per charged hadron in the final state of
$S/N_\mathrm{ch} = \SoverNchSP \pm \SoverNchSPErr$.

In the paper by Pratt and Pal the entropy was determined for the 11\% most
central Au--Au collisions at a center-of-mass energy of $\sqrt{s_\mathrm{NN}} =
\unit[130]{GeV}$. The total entropy per unit of rapidity around midrapidity was
found to be $\mathrm{d}S/\mathrm{d}y = 4451$ with an estimated uncertainty of
10\%. Using $\D N_\mathrm{ch}/\D y = 536 \pm 21$ from \cite{Alver:2010ck} and
$(\D N_\mathrm{ch}/\D y) / (\D N_\mathrm{ch}/\D \eta) \approx 1.15$ we find an
entropy per charged particle of $S/N_\mathrm{ch} \equiv (\D S/\D y)/(\D
N_\mathrm{ch}/\D y) = 7.2 \pm 0.8$. This value for Au--Au collisions at a
center-of-mass energy of $\sqrt{s_\mathrm{NN}} = \unit[130]{GeV}$ agrees with
the value of $S/N_\mathrm{ch} = \SoverNchSP \pm \SoverNchSPErr$ we obtain for
the LHC energy in this paper.

\subsection{Entropy in pp collisions at \boldmath{$\sqrt{s} = 7\,\mathrm{TeV}$}\label{sec:pp}}
Not only in high-energy nucleus-nucleus collisions but also in proton-proton
and proton-nucleus collisions transverse momentum spectra and azimuthal
distributions of produced particles can be modeled assuming a hydrodynamic
evolution of the created matter
\cite{Bozek:2016jhf,Weller:2017tsr,PHENIX:2018lia}. This provides a motivation
to determine the entropy $\D S/\D y$ with the Pal-Pratt method also in pp
collisions. Moreover, the experimentally determination of the entropy is of
interest in the context of models which are based on entropy productions
mechanisms not related to particle scatterings (see, e.g.,
\cite{Berges:2017zws,Baker:2017wtt}). Here we focus on minimum bias and 
high-multiplicity pp collisions at $\sqrt{s} = \unit[7]{TeV}$.

Transverse momentum spectra for both minimum bias collisions ($\pi$,
$\mathrm{K}$, $\mathrm{p}$ \cite{Adam:2015qaa}, $\Lambda$
\cite{Abelev:2013xaa}, and $\Xi$, $\Omega$ \cite{Abelev:2012jp}) and
high-multiplicity pp collisions ($\pi$, $\mathrm{K}$, $\mathrm{p}$
\cite{Acharya:2018orn}, $\Lambda$, $\Xi$, $\Omega$ \cite{ALICE:2017jyt}) are
taken from the ALICE experiment. The high-multiplicity sample (class~I in
\cite{ALICE:2017jyt} and \cite{Acharya:2018orn}) roughly corresponds to the
0-1\% percentile of the multiplicity distribution measured at forward and
backward pseudorapidities. HBT radii are taken from \cite{Aamodt:2011kd}. In
minimum bias pp collisions there is little dependence of $R_\mathrm{inv}$ on
transverse mass and a constant value $R_\mathrm{inv} = \unit[1.1 \pm 0.1]{fm}$
is assumed. For the high-multiplicity sample $m_\mathrm{T}$ scaling of
$R_\mathrm{inv}$ is assumed and the same power law and exponential functional
forms as in the Pb--Pb analysis are fit to the data from \cite{Aamodt:2011kd}
($N_\mathrm{ch} = 42$--$51$ class in \cite{Aamodt:2011kd}). Taking into account the
uncertainty of associating the multiplicity class in
\cite{Acharya:2018orn,ALICE:2017jyt} with the one in \cite{Aamodt:2011kd} we
assume an uncertainty of $R_\mathrm{inv}$ for the high-multiplicity sample of
about 10\%.

With
the same assumptions for the contribution of unobserved particles and feeddown
as in Pb--Pb collisions we obtain $\D S/\D y|_\mathrm{MB} = \dSdySPPPMinBias
\pm \dSdySPPPMinBiasErr$ in minimum bias (MB) collisions and $\D S/\D
y|_\mathrm{HM} = \dSdySPPPHighMult \pm \dSdySPPPHighMultErr$ for the
high-multiplicity (HM) sample. The contribution of the different particles
species to the total entropy are given in Tables~\ref{tab:dsdy_pp_mb} and
\ref{tab:dsdy_pp_hm}. 
With $\D N_\mathrm{ch}/\D\eta = 6.0 \pm 0.1$ \cite{Adam:2015gka} and $(\D
N_\mathrm{ch}/\D y) / (\D N_\mathrm{ch}/\D\eta) = 1.21 \pm 0.01$ for minimum
bias pp collisions we obtain $S/N_\mathrm{ch}|_\mathrm{MB} =
\SoverNchSPPPMinBias \pm \SoverNchSPErrPPMinBias$ for the entropy per
final-state charged particle. For the high-multiplicity sample with
$\D N_\mathrm{ch}/\D\eta = 21.3 \pm 0.6$ \cite{Acharya:2018orn} and $(\D
N_\mathrm{ch}/\D y) / (\D N_\mathrm{ch}/\D\eta) = 1.19 \pm 0.01$ we find
$S/N_\mathrm{ch}|_\mathrm{HM} = \SoverNchSPPPHighMult \pm
\SoverNchSPErrPPHighMult$.
\begin{table}[h]
\centering
\caption{\label{tab:dsdy_pp_mb}Estimate of the entropy $(\D S/\D y)_{y=0}$ in minimum bias pp collisions at
$\sqrt{s} = \unit[7]{TeV}$}
\begin{tabular}{@{}c @{\hskip 1em} c @{\hskip 1em} c @{\hskip 1em} c@{}} 
particle &  $(\D S/\D y)_{y=0}^{\text{one state}}$ & factor &  $(\D S/\D y)_{y=0}^{\text{total}}$\\
\hline \hline 
$\pi$ & 6.7 & 3 & 20.1 \\
$\mathrm{K}$ & 2.1 & 4 & 8.4 \\ 
$\eta$ & 1.4 & 1 & 1.4 \\ 
$\eta'$ &  0.3& 1 &  0.3 \\ 
$\mathrm{p}$ & 1.2 & 2 & 2.4 \\ 
$\mathrm{n}$ & 1.2 & 2 & 2.4 \\
$\Lambda$ & 0.6 & 2 & 1.2 \\ 
$\Sigma$ & 0.2 & 6 & 1.2 \\ 
$\Xi$ & 0.1 & 4 & 0.4 \\ 
$\Omega$ & 0.01 & 2 & 0.02 \\ 
\hline 
total &  &  & \dSdySPPPMinBias \\
\end{tabular}
\end{table}
\begin{table}[h]
\centering
\caption{\label{tab:dsdy_pp_hm}Estimate of the entropy $(\D S/\D y)_{y=0}$ in high-multiplicity pp collisions (class~I in \cite{ALICE:2017jyt} and \cite{Acharya:2018orn})) at
$\sqrt{s} = \unit[7]{TeV}$} 
\begin{tabular}{@{}c @{\hskip 1em} c @{\hskip 1em} c @{\hskip 1em} c@{}} 
particle &  $(\D S/\D y)_{y=0}^{\text{one state}}$ & factor &  $(\D S/\D y)_{y=0}^{\text{total}}$\\
\hline \hline 
$\pi$ & 23.8 & 3 & 71.4 \\
$\mathrm{K}$ & 7.5 & 4 & 30.0 \\ 
$\eta$ & 4.9 & 1 & 4.9 \\ 
$\eta'$ &  1.0 & 1 &  1.0 \\ 
$\mathrm{p}$ & 4.2 & 2 & 8.4 \\ 
$\mathrm{n}$ & 4.2 & 2 & 8.4 \\
$\Lambda$ & 2.3 & 2 & 4.6 \\ 
$\Sigma$ & 0.8 & 6 & 4.8 \\ 
$\Xi$ & 0.5 & 4 & 2.0 \\ 
$\Omega$ & 0.1 & 2 & 0.2 \\ 
\hline 
total &  &  & \dSdySPPPHighMult \\
\end{tabular}
\end{table}

\section{Comparisons to statistical models\label{sec:hrg}} 
In order to compare the $S/N_\mathrm{ch}$ value determined from the measured final
state particle spectra to calculations in which particles originate from a hadron
resonance gas one needs to know the ratio $N/N_\mathrm{ch}$ of the total
number of primary hadrons $N (\equiv N_\mathrm{prim})$ to the total number of
measured charged hadrons in the final state $N_\mathrm{ch} (\equiv
N_\mathrm{ch}^\mathrm{final})$. The latter contains feed-down contributions
from strong and electromagnetic hadron decays. If only pions were produced one
would get $N/N_\mathrm{ch} = 3/2$. With the aforementioned Pythia 8.2
simulation and the list of stable hadrons implemented in Pythia (again with
hadron yields given by \Eq{eq:particle_density} for $T = \unit[156]{MeV}$
and $\mu_\mathrm{b} = 0$) we obtain a value of
($N/N_\mathrm{ch})_\mathrm{Pythia} = 1.14$. In this calculation particles with
a lifetime $\tau$ above $\unit[1]{mm}/c$ were considered stable. Using the
implementation of the hadron resonance gas of ref.~\cite{Vovchenko:2019pjl} we
find ($N/N_\mathrm{ch})_\mathrm{TF} = 1.09$. In following we use
$N/N_\mathrm{ch} = \NoverNch \pm \NoverNchErr$, i.e., we take the average of
the two results as central value and the difference as a measure of the
uncertainty.

\begin{figure}
\includegraphics[width=0.95\linewidth]{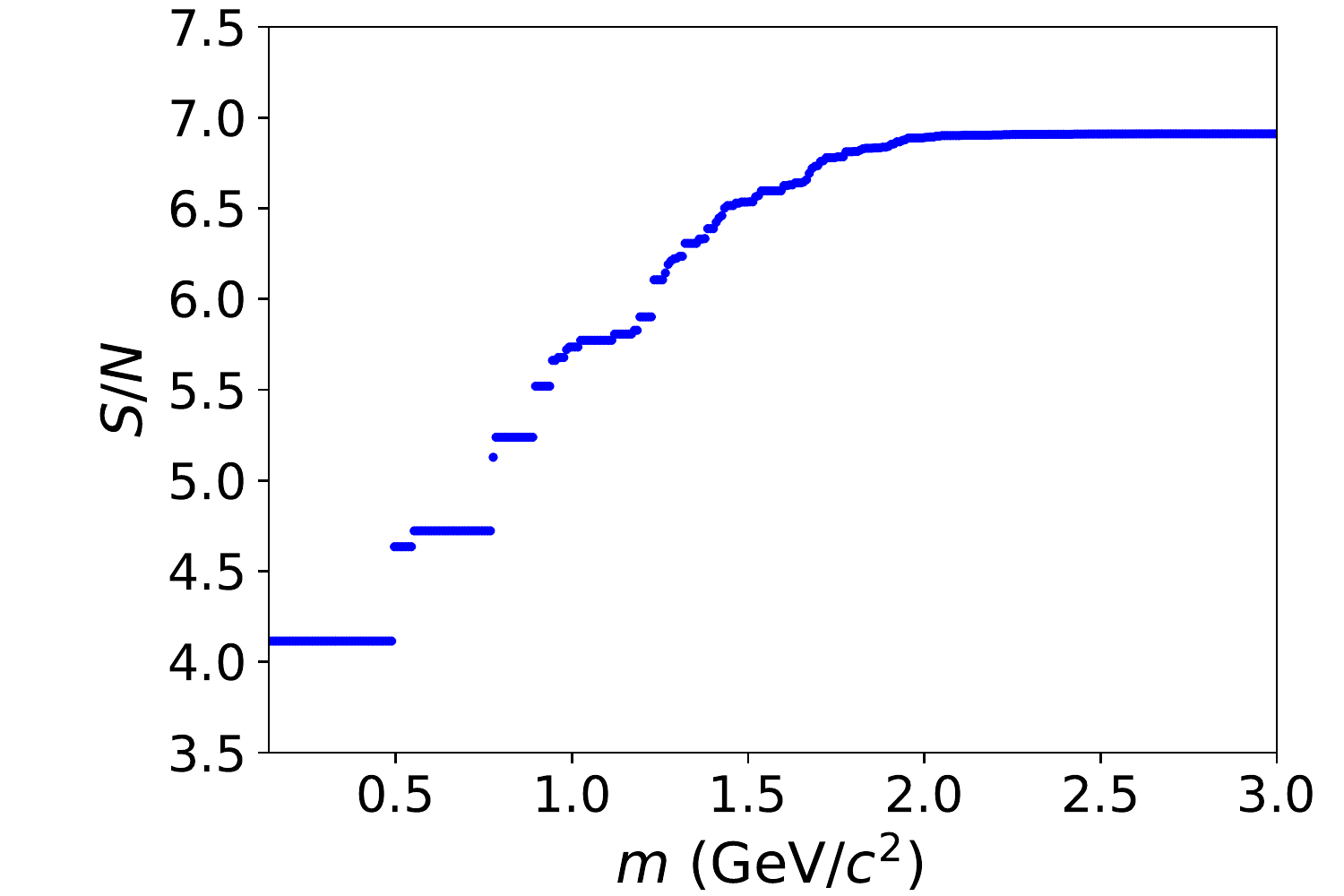}
\caption{Entropy per primary hadron $S/N$ for a non-interacting thermal hadron resonance
gas at a temperature of $T = \unit[156]{MeV}$ as given by
\Eq{eq:s_over_n} as a function of the upper mass limit for particles
listed in the particle data book \cite{Tanabashi:2018oca}. The entropy per
hadron saturates for high upper mass limits at a value of $S/N = 6.9$.}
\label{fig:s_over_n_vs_mass_cut} 
\end{figure}

In the simplest form of the description of a hadron resonance gas the system is
treated as a non-interacting gas of point-like hadrons where hadronic
resonances have zero width. The entropy density for a primary hadron with mass $m$ at
thermal equilibrium with temperature $T$ and vanishing chemical potential $\mu = 0$ is then given by~
\cite{Letessier:2002gp}
\begin{multline}
s = \frac{4 g T^3}{2 \pi^2} \sum_{k=1}^{\infty} \frac{(\pm)^{k+1}}{k^4} 
\left[\left(\frac{km}{T}\right)^2 \mathrm{K}_2\left(\frac{km}{T}\right) +  \right. \\ 
\left. \frac{1}{4} \left(\frac{km}{T}\right)^3 \mathrm{K}_1\left(\frac{km}{T} \right)\right]
\label{eq:s_hrg}
\end{multline}
where $+$ is for bosons and $-$ for fermions. $\mathrm{K}_1$ and $\mathrm{K}_2$ are 
modified Bessel functions of the second kind.
Using Eqs.~(\ref{eq:particle_density}) and (\ref{eq:s_hrg}) the
entropy per primary hadron in the thermal hadron resonance gas can be calculated as
\begin{equation}
S/N = \frac{\sum_i s_i}{\sum_i n_i}
\label{eq:s_over_n}
\end{equation}
where the index $i$ denotes the different particles species.
For illustration, the entropy per hadron is shown in
Figure~\ref{fig:s_over_n_vs_mass_cut} as a function of the upper limit on the
mass for all particles listed in the particle data book
\cite{Tanabashi:2018oca}.  

More sophisticated implementations of the hadron resonance gas take the volume
of the hadrons and the finite width of hadronic resonances into account
\cite{Andronic:2017pug,BraunMunzinger:2001mh,Andronic:2008gu,Torrieri:2004zz,
Torrieri:2006xi,Petran:2013dva,Wheaton:2004qb,Kisiel:2005hn,Chojnacki:2011hb,
Vovchenko:2019pjl}.
Some of these models implement chemical non-equilibrium factors which we do not
consider here. Models can also differ in the set of considered hadron states. In
the following we concentrate on the models by Braun-Munzinger et al.\
\cite{Andronic:2017pug} (``model 1'') and Vovchenko/St\"{o}cker \cite{Vovchenko:2019pjl} 
(``model 2'').
The corresponding values for the entropy per primary hadron $S/N$ and the
entropy per final state charged hadron $S/N_\mathrm{ch}$ are given in
Tab.~\ref{tab:stat_model}. The $S/N_\mathrm{ch}$ values for these models are
somewhat larger than the measured value of $S/N_\mathrm{ch} = \SoverNchSP \pm
\SoverNchSPErr$, but the deviations are not larger than 1--2$\sigma$. We note here
that the two approaches calculate slightly different quantities. Our estimate
is based on the non-equilibrium distributions of a few final state hadrons, while
\Eq{eq:s_over_n} sums the entropy contributions of all primary hadrons in a thermal state before
the decays.
Although on general grounds we expect the total entropy to increase
during the 
decays and re-scatterings in the hadronic phase, there are some decay products, e.g.\ photons,
which are not included in our current entropy count. Accounting for such differences
between the two approaches might
bring the estimates closer together.
\begin{table} 
\caption{\label{tab:stat_model} Entropy per primary hadron $S/N$ at a
temperature of $T = \unit[156]{MeV}$ for different hadron resonance gas models. 
The entropy per final state charged hadron is
calculated from $S/N$ by multiplying with the factor $N/N_\mathrm{ch} =
\NoverNch \pm \NoverNchErr$. The volume correction of model 2 is the based on
the Quantum van der Waals model. Within 1--2$\sigma$ the $S/N_\mathrm{ch}$
values of these models agree with the value of $S/N_\mathrm{ch} = \SoverNchSP \pm
\SoverNchSPErr$ obtained from data.} 
\begin{tabular}{@{}l @{\hskip 1em} c@{\hskip 1em} c @{}} Model & $S/N$ & $S/N_\mathrm{ch}$ \\ 
\hline \hline
Simple HRG (\Eq{eq:s_over_n}) & 6.9 & $7.7 \pm 0.2$ \\
\hline 
Model 1 (Braun-Munzinger et al. \cite{Andronic:2017pug, Andronic:2012ut})& & \\ 
without volume correction & 7.3 & $8.1 \pm 0.2$ \\ 
with volume correction & 7.6 & $8.5 \pm 0.2$ \\ 
\hline 
Model 2 (Vovchenko, St\"ocker \cite{Vovchenko:2019pjl})& & \\
ideal & 6.9 & $7.7 \pm 0.2$ \\ 
with volume correction, zero width & 7.2 & $8.1 \pm 0.2$ \\
with volume correction, finite width & 7.1 & $7.9 \pm 0.2$ \\ 
\end{tabular}
\end{table}

\section{Initial conditions and entropy production}
\subsection{Pb--Pb collisions\label{sec:IC}} 

The entropy in nuclear collisions, which we calculated in previous sections, is
not created instantaneously, but rather the entropy production takes place in
several stages in nuclear collisions~\cite{Muller:2011ra}. In this section we will 
use different models to describe the boost invariant expansion and, in particular,
to determine the average initial conditions in  0--10\% most central Pb--Pb collisions at $\sqrt{s_\mathrm{NN}} = \unit[2.76]{TeV}$
at time $\tau_0=\unit[1]{fm}$.

First, we can make an estimate of the initial 
temperature $T(\tau_0)$ by simplifying the early time evolution of the QGP.
As it was done in the original work by Bjorken~\cite{Bjorken:1982qr},
we will consider a one-dimensional boost-invariant expansion of homogeneous 
plasma with the transverse extent determined
by the geometry of the nuclei, i.e.\
the transverse area $A$.
The time evolution of the energy density is then solely a function of proper time $\tau$ and
the assumed constituent equation for the longitudinal pressure $P_L= \tau^2 T^{\eta\eta}$.
For collisionless gas with $P_L=0$, the energy density falls as  $e\propto\tau^{-1}$,
i.e.\ energy per rapidity $\D E/\D y = e\tau A$ is constant~\cite{Bjorken:1982qr}. For an ideal fluid with
equation of state $p= c_s^2 e$, where  $c_s$ is the (constant) speed of sound, the
energy density decreases faster, $e \propto \tau^{-1-c_s^2}$, because of the work done against the longitudinal pressure~\cite{Bjorken:1982qr,Gyulassy:1983ub, Danielewicz:1984ww}. However, total entropy per rapidity  $\D S/\D y = s\tau A$ 
stays constant irrespective of the equation of state as long as viscous dissipation can be
neglected~\cite{Bjorken:1982qr,Gyulassy:1983ub,Danielewicz:1984ww, Hwa:1985xg}.
Below we compare the initial temperature estimates at time $\tau_0=\unit[1]{fm}$ obtained from the final
entropy $\D S/\D y $ and energy $\D E/\D y $
using, respectively, isentropic ideal fluid and free-streaming evolutions of QGP%
~\footnote{For initial energy estimates at even earlier times, see a recent publication~\cite{Giacalone:2019ldn}, where a generalised constitutive equation of a hydrodynamic
attractor was considered.}.

 Assuming that the subsequent near ideal
hydrodynamic evolution does not change the total entropy per rapidity ${\D
S}/{\D y}$ (which  is also true for free-streaming expansion)
 the initial entropy density is equal to
\begin{equation} 
  s(\tau_0) = \left.\frac{1}{A\tau_0} \frac{\D S}{\D y}\right|_{y=0}
.
\label{eq:entropy_density_from_dsdy}
\end{equation} 
Using for the transverse area $A =
\pi R_\mathrm{Pb}^2$ with $R_\mathrm{Pb} = \unit[6.62]{fm}$ \cite{DeJager:1987qc,Abelev:2013qoq} 
gives an initial entropy density
for the 0--10\% most central Pb--Pb collisions at $\sqrt{s_\mathrm{NN}} = \unit[2.76]{TeV}$
\begin{equation}
  s(\tau_0) = \unit[82.3]{fm^{-3}}\label{eq:naiveS}
.\end{equation}
According to the lattice QCD equation of state~\cite{Bazavov:2014pvz,Borsanyi:2016ksw}, this corresponds to a 
temperature 
\begin{equation}
T(\tau_0) \approx \unit[340]{MeV} 
.\end{equation}

The transverse energy at midrapidity for the 10\% most Pb--Pb
collisions at $\sqrt{s_\mathrm{NN}} = \unit[2.76]{TeV}$ was measured to be $\D
E_\mathrm{T}/\D y \approx
\unit[1910]{GeV}$ \cite{Chatrchyan:2012mb,Adam:2016thv}. Using again $A = \pi
R_\mathrm{Pb}^2$ with $R_\mathrm{Pb} = \unit[6.62]{fm}$ as an approximation
for the transverse overlap area the initial energy density can be calculated
according to the Bjorken formula~\cite{Bjorken:1982qr}
\begin{equation}
 e(\tau_0)= \frac{1}{A\tau_0} 
 \left. \frac{\D E_\mathrm{T}}{\D y} \right|_{y=0},\label{eq:Bjorken}
\end{equation}
which gives an energy density $e(\tau_0) \approx
\unit[13.9]{GeV/fm^3}$. This would correspond to much lower initial temperature $T(\tau_0)\approx\unit[305]{MeV}$.
This is because \Eq{eq:Bjorken} is derived under the assumption of a constant
energy per rapidity~\cite{Bjorken:1982qr}. This holds for a free-streaming (or
pressureless) expansion, but in hydrodynamics the system cools down faster due
to work done against the longitudinal pressure. Taking $\tau_\text{f} =
R_\text{Pb}$ as a rough estimate for the lifetime of the fireball, ideal
hydrodynamics predicts an $\left( \tau_{\text{f}}/\tau_0
\right)^{\frac{1}{3}}\approx 1.9 $ times larger initial energy density, which
would revise the initial temperature estimate upwards to
$T(\tau_0)\approx\unit[355]{MeV}$ and closer to the value we obtained from the
entropy method.

Instead of assuming a constant entropy density in a collision, it is more
realistic to use an entropy density profile $s(\tau, \vec r)$, where $\vec r$
is a two-dimensional vector in the transverse plane (we still assume
boost-invariance in the longitudinal direction). We will employ the
two-component optical Glauber model to construct the transverse profile of
initial entropy density~\cite{Miller:2007ri}. In this model the initial
entropy is proportional to the participant nucleon number and the number of
binary collisions. For a collision at impact parameter $\vec b$, the entropy
profile is then
\begin{equation}
  s(\tau_i,\vec{r}; \vec b)  =\frac{\kappa_s}{\tau_i} \left(\frac{1-\alpha}{2} \frac{\D N^\text{part}(\vec r, \vec b)}{\D^2r}
  + \alpha \frac{\D N^\text{coll}(\vec r, \vec b)}{\D^2r}\right), \label{eq:Glauber}
\end{equation}
where $\kappa_s(1-\alpha)/2$ is entropy per rapidity per participant and
$\kappa_s \alpha$ is entropy per rapidity per binary collision.  The number
densities are calculated using the nucleon-nucleon thickness functions  (see
\app{sec:Glauber} for details), and the value  $\alpha=0.128$ 
reproduces centrality dependence of multiplicity~\cite{Abelev:2013qoq}.
We average over the impact parameter $|\vec b|\leq \unit[4.94]{fm}$ 
to produce entropy profile corresponding to 0-10\% centrality bin of Pb--Pb collisions at
$\sqrt{s_\mathrm{NN}} = \unit[2.76]{TeV}$
~\cite{Abelev:2013qoq}.
The overall normalization factor $\kappa_s$ is adjusted to reproduce the final
state entropy estimated in \Sec{sec:PbPb}, which depends on the expansion model. 

\begin{figure}
\centering
\includegraphics[width=\linewidth]{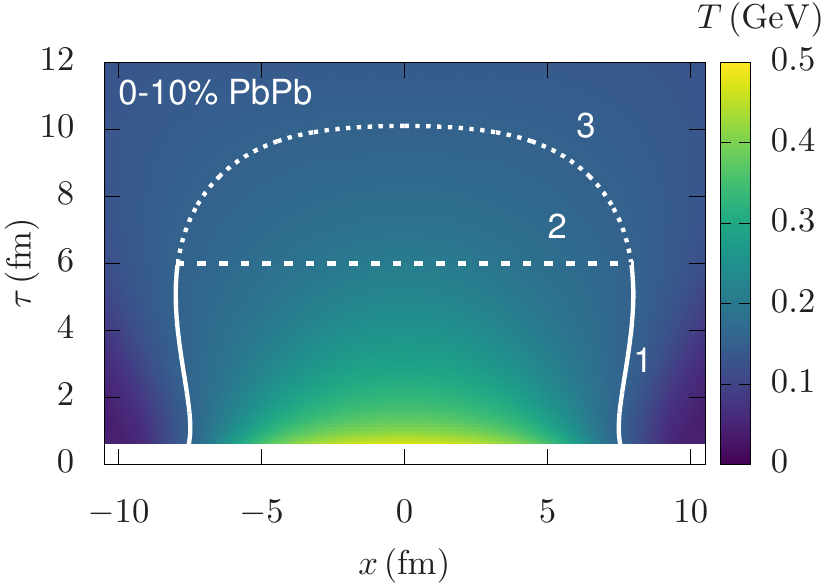}
\caption{Temperature in hydrodynamic evolution of an averaged 0--10\% Pb--Pb event at $\sqrt{s_\mathrm{NN}} = \unit[2.76]{TeV}$. Lines 1 and 3 are the freeze-out 
  $T_\text{fo}=\unit[156]{MeV}$ contour, whereas line 2 indicates a constant
  time contour in the QGP phase, i.e. $T(\tau,\vec{r})>T_\text{fo}$.
Initial conditions for hydrodynamics at $\tau_\text{hydro}=\unit[0.6]{fm}$ were provided
by \kompost{} pre-equilibrium evolution from the starting time of $\tau_\text{EKT}=\unit[0.1]{fm}$.}
\label{fig:Tcon}
\end{figure}
To simulate the evolution and entropy production in nucleus-nucleus collisions
we employ two recently developed models:
kinetic pre-equilibrium propagator \kompost{}~\cite{Kurkela:2018wud,Kurkela:2018vqr},
and viscous relativistic hydrodynamics code FluiduM~\cite{Floerchinger:2018pje}\footnote{We neglect the entropy production in the hadronic phase and match the entropy on the freeze-out surface.}.
For simplicity we employ a constant value 
of specific shear viscosity $\eta/s$ and vanishing bulk viscosity $\zeta/s$ throughout the evolution.

 \kompost{} uses  linear response functions obtained from QCD kinetic
 theory\footnote{The current implementation of \kompost{} uses results of pure
 glue simulations, but recent calculations with full QCD degrees of freedom
 indicate that the evolution of the total energy-momentum tensor will not be
 significantly altered  by chemical
 equilibration~\cite{Kurkela:2018oqw,Kurkela:2018xxd}.} to propagate and
 equilibrate the highly anisotropic initial energy momentum tensor, which can
 be specified at an early starting time $\tau_\text{EKT}=\unit[0.1]{fm}$. We
 specify the initial energy-momentum tensor profile to be\footnote{As a purely
 practical tool we use a lattice equation of state to convert entropy density
 profile obtained from the Glauber model \Eq{eq:Glauber} to the energy density
 needed to initialize \kompost{}, even though the system at
 $\tau_\text{EKT}=\unit[0.1]{fm}$ is not in thermodynamic equilibrium. }
\begin{equation}
T^{\mu \nu}(\tau_\text{EKT},\vec r) = e(\tau_\text{EKT},\vec r)\, \text{diag}\,\left(1, \tfrac{1}{2}, \tfrac{1}{2},0\right)
.\end{equation}
At the end of \kompost{} evolution all components of the energy momentum tensor, 
the energy density, transverse flow and the
shear-stress components, are passed to the hydrodynamic model at fixed time $\tau_\text{hydro}=\unit[0.6]{fm}$.

The FluiduM package solves the linearized Israel-Stewart type hydrodynamic
equations around an azimuthally symmetric background profile. In this work we
propagate the radial background profile until the freeze-out condition is met,
which we define by a constant freeze-out temperature
$T_\text{fo}=\unit[156]{MeV}$. Above this temperature the equation of state is
that of lattice QCD~\cite{Borsanyi:2016ksw}. Unless otherwise stated, we use a
constant specific shear viscosity $\eta/s=0.08$ and vanishing bulk viscosity
$\zeta/s=0$.

\begin{figure*}
\centering
\subfig{a}{\includegraphics[width=0.45\linewidth]{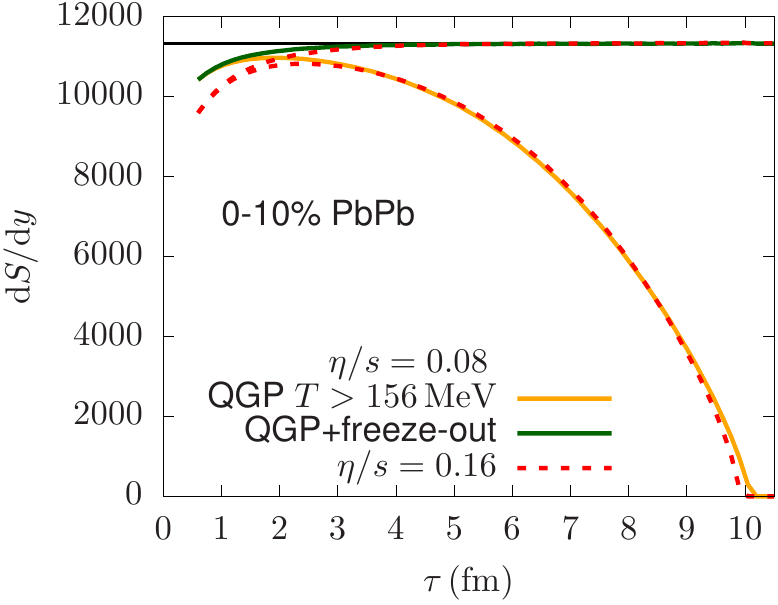}}\quad
\subfig{b}{\includegraphics[width=0.45\linewidth]{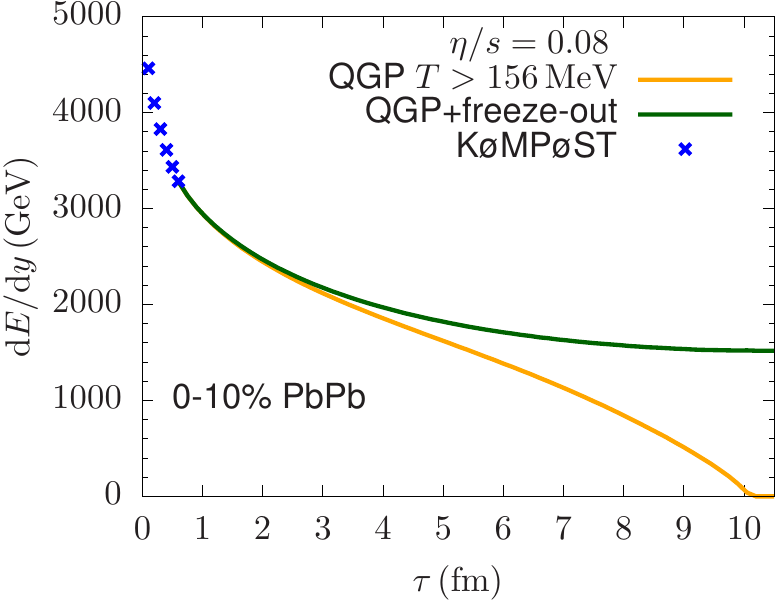}}
\caption{(a) Entropy per rapidity in viscous hydrodynamic expansion 
  with specific shear viscosity $\eta/s=0.08$ for central $\sqrt{s_\mathrm{NN}} =
\unit[2.76]{TeV}$ Pb--Pb collisions (centrality class 0-10\%) as a function of time of
  contour $2$ in \Fig{fig:Tcon}.
 The yellow
  line corresponds to entropy in the QGP phase ($T(\tau,r)>T_\text{fo}$)
  (contour $2$ in \Fig{fig:Tcon}), whereas the green line shows the total
  cumulative entropy (contour $1+2$ in \Fig{fig:Tcon}).  Dashed red lines show
  the corresponding result for a simulation with $\eta /s=0.16$. In both cases
the initial conditions, i.e.\ parameter $\kappa_s$ in \Eq{eq:Glauber}, were
tuned to reproduce the final
freeze-out entropy $\D S /\D y=\dSdySP$ after the pre-equilibrium and hydrodynamic evolutions. (b) Analogous plot for energy per rapidity in 
hydrodynamic expansion with $\eta /s=0.08$. The additional points show 
energy per rapidity in the pre-equilibrium stage.  }
\label{fig:dSdy}
\end{figure*}

We start by showing the temperature evolution in the hydrodynamic phase in \Fig{fig:Tcon}. The combined
solid and dotted white lines represent the freeze-out line at
$T_\text{fo}=\unit[156]{MeV}$. The dashed horizontal line indicates the spatial slice of the
fireball at some fixed time $\tau$ and above the freeze-out temperature.
We now can define entropy as an integral of the entropy current $su^\mu$ over a hypersurface $\Sigma_i$
where $\Sigma_i$ is one or more of the contours shown in \Fig{fig:Tcon}.
We define the total entropy in the QGP state at time $\tau$ as the integral over the contour 2: 
\begin{align}
  \left.S(\tau)\right|_{\text{QGP}} \equiv \int_{\Sigma_2(\tau)}  \D\sigma_\mu su^\mu.
\end{align}
To include the entropy outflow from the QGP because of freeze-out we also define entropy
on the contours $\Sigma_1(\tau'<\tau)+\Sigma_2(\tau)$:
\begin{align}
  \left.S(\tau)\right|_{\text{QGP+freeze-out}} \equiv \int_{\Sigma_1(\tau'<\tau)+\Sigma_2(\tau)}  \D\sigma_\mu su^\mu.
\end{align}
Because of viscous dissipation $\left.S(\tau)\right|_{\text{QGP+freeze-out}}$ increases until the temperature in every hydro cell drops below the freeze-out temperature
  and the maximum value is simply the entropy current integral over the freeze-out surface  $\Sigma_1(\tau'<\tau)+\Sigma_3(\tau'>\tau)$.
In \Fig{fig:dSdy}(a)  we show the time dependence of entropy per rapidity in
    the QGP phase (yellow line) and including freeze-out outflow (green line)
    in hydrodynamically expanding plasma.  The solid lines are for the
    simulation with
 with $\eta/s=0.08$  and dashed lines correspond to $\eta/s=0.16$.
In both cases
the initial entropy profile, \Eq{eq:Glauber}, is adjusted so that
after the pre-equilibrium (\kompost{}) and hydrodynamic (FluiduM) 
evolution the final
entropy on the freeze-out surface is equal to $\D S /\D y=\dSdySP$ estimated in
\Sec{sec:PbPb}.
We see that at early times  entropy is produced rapidly, but there is little
entropy outflow through the freeze-out surface. At $\tau\approx\unit[2]{fm}$
the entropy in the hot QGP phase starts to drop because matter is crossing
the freeze-out surface and at  $\tau\approx
\unit[10]{fm}$ there is no hot QGP phase left.

Here we note that the early time viscous entropy production in the
hydrodynamic phase depends strongly on the initialization of the shear-stress
tensor. In this work we use the pre-equilibrium propagator \kompost{}, which
provides all components of energy-momentum tensor at hydrodynamic starting
time and the shear-stress tensor approximately satisfies the Navier-Strokes
constitutive equations~\cite{Kurkela:2018wud,Kurkela:2018vqr}. We determine
that for evolution with $\eta /s=0.08$ the entropy per rapidity at time
$\tau_0=\unit[1.0]{fm}$  is $\approx 95\%$ of the final entropy on the
freeze-out. For twice larger shear viscosity the entropy production doubles
and to produce the same final entropy we need only $\approx 90\%$ at
$\tau_0=\unit[1.0]{fm}$. Such entropy production is neglected in the naive
estimate of \Eq{eq:entropy_density_from_dsdy}.

\begin{figure*}
\centering
\subfig{a}{\includegraphics[width=0.45\linewidth]{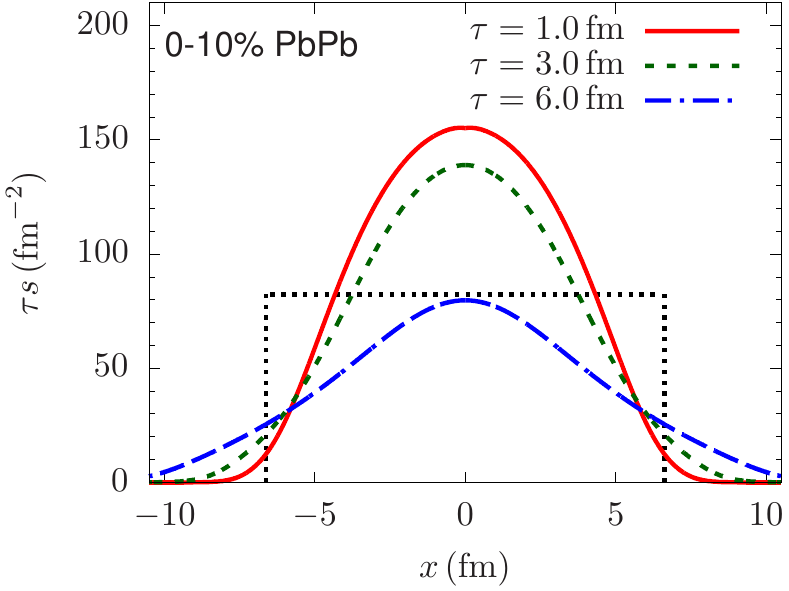}}\quad
\subfig{b}{\includegraphics[width=0.45\linewidth]{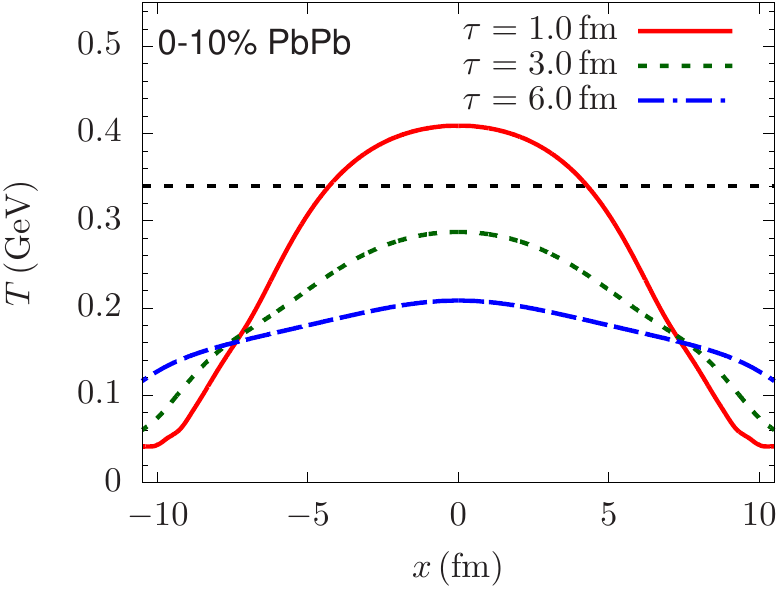}}
\caption{(a) Entropy density profile (multiplied by $\tau$) in viscous
  hydrodynamic simulation with $\eta/s=0.08$ at times $\tau=\unit[1,3,6]{fm}$. The black dotted
  square indicates the initial entropy density estimate $\tau_0 s(\tau_0) = \unit[82.3]{fm}^{-2}$, see
\Eq{eq:entropy_density_from_dsdy}. (b) Temperature profile at $\tau=\unit[1,3,6]{fm}$.
The black dotted line corresponds to $T = \unit[340]{MeV}$.}
\label{fig:entrslice}
\end{figure*}

Analogously to entropy, we use the same contours to define energy in the
collision, that is, as integrals of the energy current $e u^\mu$. In
\Fig{fig:dSdy}(b) we show the energy per rapidity in different phases of the
collision. We confirm that the energy per rapidity decreases rapidly in the
hydrodynamic phase and at $\tau_0=\unit[1.0]{fm}$ is nearly twice larger than
on the freeze-out surface and therefore invalidating the naive initial energy
density estimates using Bjorken formula \Eq{eq:Bjorken}. However we do note
that the magnitude of the final energy per rapidity in our event is below the
measured value. In addition we show points for the energy per rapidity in the
pre-hydro phase simulated by \kompost{}. Despite the large anisotropy in the
initial energy-momentum tensor ($T^{zz}\approx 0$ initially), the energy per
rapidity is rapidly decreasing in this phase. We note that at the same time
there is a significant entropy production in the kinetic pre-equilibrium
evolution~\cite{Kurkela:2018vqr}.

Next in \Fig{fig:entrslice}(a) we look at the transverse entropy density
profile $\tau s(\tau,\vec{r})$ at different times
$\tau=\unit[1.0,3.0,6.0]{fm}$ in the hydrodynamic evolution with
$\eta/s=0.08$. We see that the profile changes only little between
\unit[1]{fm} and
\unit[3]{fm}, which is because of an approximate one dimensional expansion and
viscous entropy production. At later times the profile expands radially and
drops in magnitude.  The black-dotted line indicates the naive estimate of
entropy density $\tau_0 s(\tau_0)=\unit[82.3]{fm}^{-2}$ for a disk-like
profile with radius $R_\text{Pb}=\unit[6.62]{fm}$, see
\Eq{eq:naiveS}. Despite an over-estimation of the net entropy at
$\tau_0=\unit[1]{fm}$, the actual density at the center of entropy profile is
twice larger than the naive estimate. Correspondingly, the transverse
temperature profile at $\tau_0=\unit[1]{fm}$, shown in
\Fig{fig:entrslice}(b), is larger than the simple estimate and can reach
$\unit[400]{MeV}$ in the center of the fireball.

\subsection{Central pp collisions\label{sec:ICpp}} 
\begin{figure}
\centering
\includegraphics[width=\linewidth]{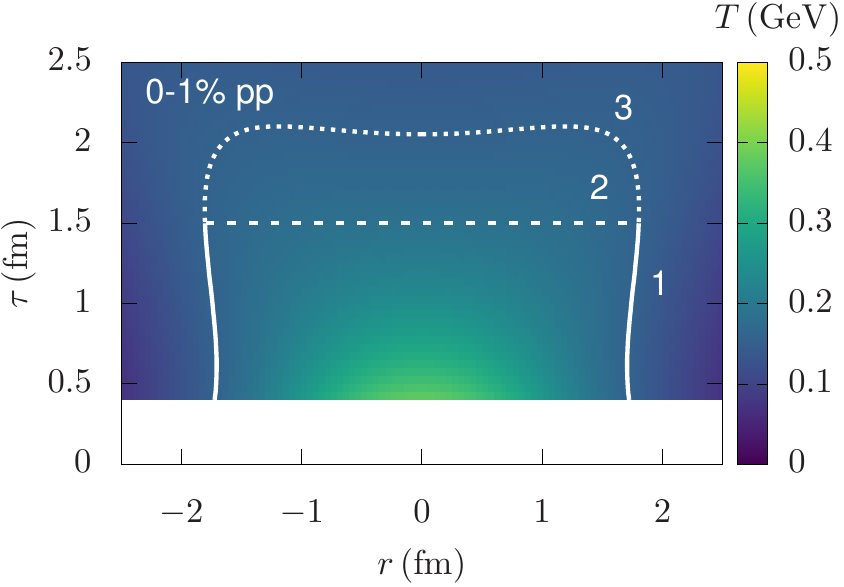}
\caption{Temperature in hydrodynamic evolution of an averaged 0--1\% pp
  event at $\sqrt{s} = \unit[7]{TeV}$. Lines 1 and 3 are the freeze-out
  $T_\text{fo}=\unit[156]{MeV}$ contour, whereas line 2 indicates a constant
  time contour in the QGP phase, i.e. $T(\tau,\vec{r})>T_\text{fo}$. Initial
  conditions for hydrodynamics at $\tau_\text{hydro}=\unit[0.4]{fm}$ were
  provided by \kompost{} pre-equilibrium evolution from the starting time of
  $\tau_\text{EKT}=\unit[0.1]{fm}$.}
\label{fig:ppTcon}
\end{figure}

In this section we present a similar analysis of entropy production in
ultra-central pp collisions. Because of much smaller initial size, the QGP
fireball (if created),  has a much shorter life-time than the central Pb--Pb
collisions. This should enhance the relative role of the pre-equilibrium
physics of QGP formation.

To model the initial entropy density in pp collision, we use a Gaussian parametrization
of the transverse entropy distribution
\begin{equation}
s(\tau_0,\vec r) = \frac{\kappa_s}{\tau_0 2\pi \sigma^2} e^{-\frac{r^2}{2\sigma^2}}
\label{eq:Gaus}
\end{equation}
with a width $\sigma = \unit[0.6]{fm}$, as used in other
parametrizations~\cite{Moreland:2014oya}. We use a fixed value of
$\eta/s=0.08$ and, in view of the range of applicability of the linearized
pre-equilibrium propagator, we use \kompost{} for a shorter time from
$\tau_\text{EKT}=\unit[0.1]{fm}$ to $\tau_\text{hydro}=\unit[0.4]{fm}$.

\begin{figure*}
\centering
\subfig{a}{\includegraphics[width=0.45\linewidth]{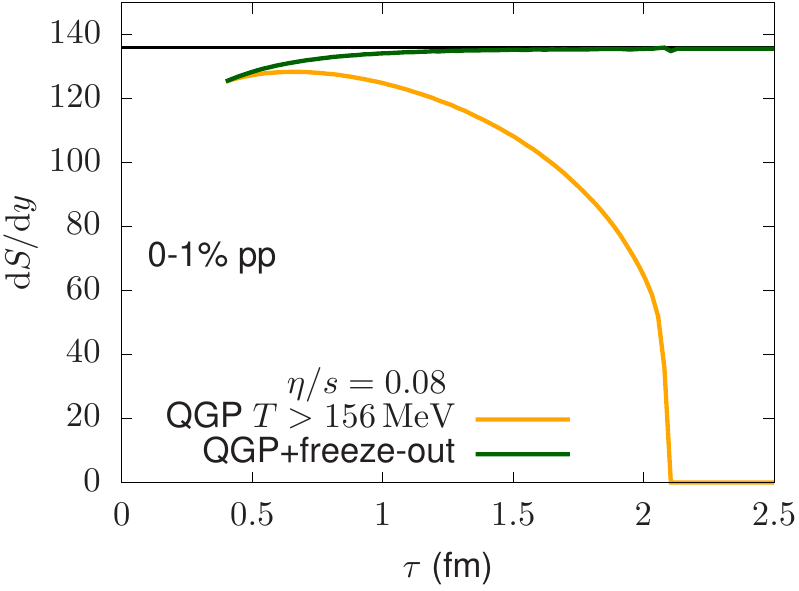}}\quad
\subfig{b}{\includegraphics[width=0.45\linewidth]{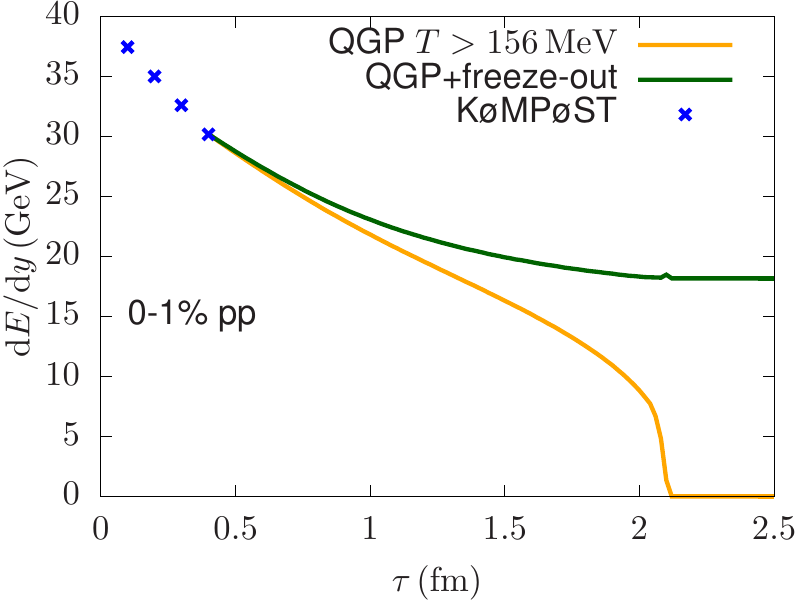}}
\caption{(a) Entropy per rapidity in viscous hydrodynamic expansion 
  with specific shear viscosity $\eta/s=0.08$ for $\sqrt{s} = \unit[7]{TeV}$
  pp collisions (0--1\% collisions with the highest multiplicity)
as a function of time of
  contour $2$ in \Fig{fig:ppTcon}.
  The
  yellow line corresponds to entropy in the QGP phase
  ($T(\tau,r)>T_\text{fo}$) (contour $2$ in
  \Fig{fig:ppTcon}), whereas the green line shows the total cumulative entropy
  (contour $1+2$ in \Fig{fig:ppTcon}). The initial conditions, i.e.\ parameter
  $\kappa_s$ in \Eq{eq:Gaus}, was tuned to reproduce the final freeze-out
  entropy $\D S /\D y=\dSdySPPPHighMult$ after the pre-equilibrium and
  hydrodynamic evolutions. (b) Analogous plot for energy per rapidity in
  hydrodynamic expansion with $\eta /s=0.08$. Additional points show energy
  per rapidity in the pre-equilibrium stage.  }
\label{fig:dSdypp}
\end{figure*}

First we show the temperature evolution in \Fig{fig:ppTcon} and indicate the
freeze-out contour (lines 1 and 3). We note that because of the compact
initial size, the transverse expansion is so explosive that the center of the
fireball actually freezes-out before the edges (similar results were found in
Refs.~\cite{Niemi:2014wta, Heinz:2019dbd}). Next in \Fig{fig:dSdypp}(a) we
show the entropy evolution in the QGP phase and together with the outflow from
through the freeze-out surface. In a smaller system, the radial flow builds up
faster and the QGP and the combined QGP+freeze-out surface contributions
starts to deviate early. This does not capture the entropy which already left
$T>T_\text{fo}$ region in the \kompost{} phase, but for the early hydro
starting time $\tau_\text{hydro}$, this fraction is small. We see that as a
fireball of QGP ultra central pp collisions have a lifetime just above
$\tau=\unit[2]{fm}$. Therefore the $\tau_0=\unit[1]{fm}$ reference time is no
longer adequate time to discuss the ``initial conditions'' in such collisions.
Next, in \Fig{fig:dSdypp}(b) we show the energy per rapidity in the
hydrodynamic and pre-equilibrium stages. Here again we see that energy per
rapidity decreases more rapidly in comparison of entropy production.

\begin{figure*}
\centering
\subfig{a}{\includegraphics[width=0.45\linewidth]{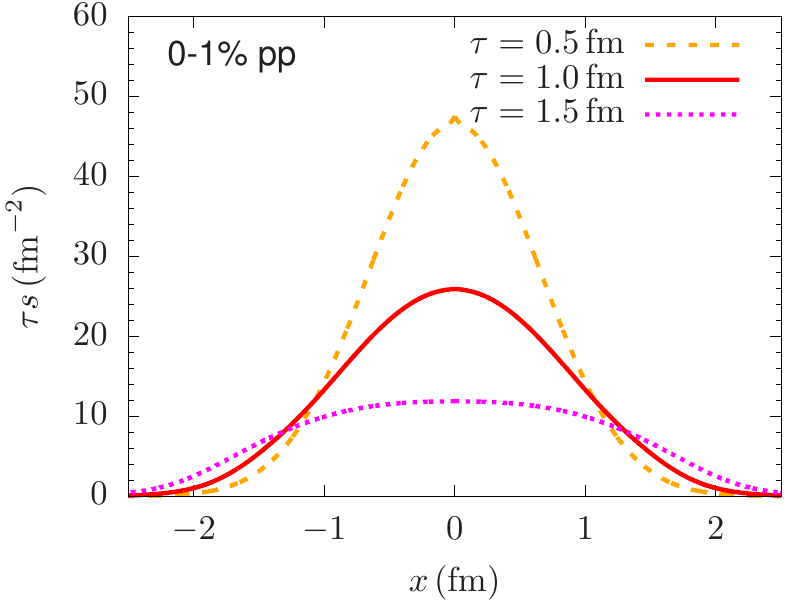}}\quad
  \subfig{b}{\includegraphics[width=0.45\linewidth]{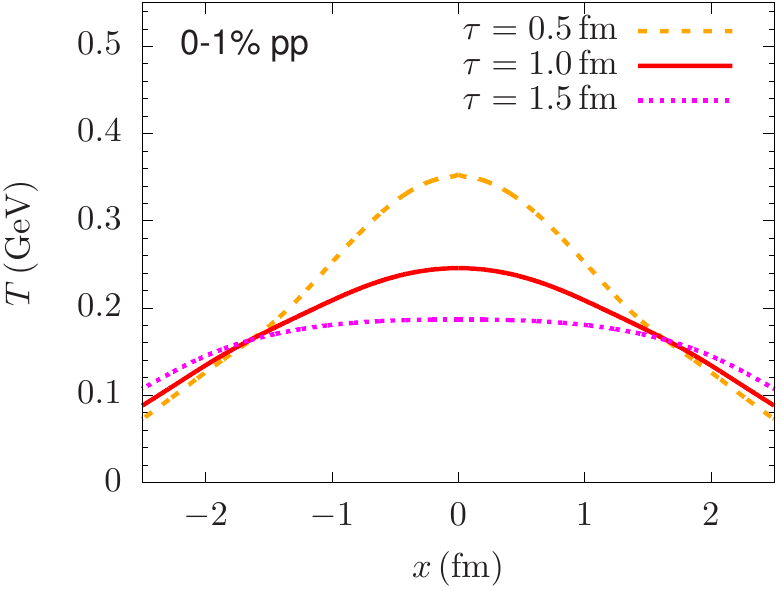}}
\caption{(a) Entropy density profile (multiplied by $\tau$) in viscous
hydrodynamic simulation with $\eta/s=0.08$ at times $\tau=\unit[0.5,1.0,1.5]{fm}$. 
(b) Corresponding temperature profiles at $\tau=\unit[0.5,1.0,1.5]{fm}$.}
\label{fig:entrslicepp}
\end{figure*}

For the transversely resolved picture of entropy and temperature profiles, we
supply figures \Figs{fig:entrslicepp}(a) and \ref{fig:entrslicepp}(b)
correspondingly. At $\tau_0=\unit[1]{fm}$ the maximum entropy density is much
smaller than in 0-10\% centrality Pb--Pb collisions and only at $\tau=\unit[0.5]{fm}$
the temperature at the center reaches above  $T=\unit[300]{MeV}$.

\section{Summary and Conclusions\label{sec:summary}}
We provide independent determination of the final-state entropy $\D S/\D y$ in $\sqrt{s} =
\unit[7]{TeV}$ pp and $\sqrt{s_\mathrm{NN}} =
\unit[2.76]{TeV}$ Pb--Pb collisions from the final phase space density calculated from the experimental data of identified particle spectra and HBT radii. In addition, we have calculated the entropy
per final-state charged hadron $(\D S/\D y) / (\D N_\mathrm{ch}/\D y)$ in different collision systems. We
find the following values for pp and Pb--Pb collisions:
\begin{table}[h!]
\begin{tabular}{ccc}
system & $\D S/\D y$ & $(\D S/\D y) / (\D N_\mathrm{ch}/\D y)$ \\
\hline 
Pb--Pb, 0--10\%& $\dSdySP \pm \dSdySPErr$ & $\SoverNchSP \pm \SoverNchSPErr$ \\
pp minimum bias &  $\dSdySPPPMinBias \pm \dSdySPPPMinBiasErr$ & $\SoverNchSPPPMinBias \pm \SoverNchSPErrPPMinBias$ \\
pp high mult. & $\dSdySPPPHighMult \pm \dSdySPPPHighMultErr$ & $\SoverNchSPPPHighMult \pm \SoverNchSPErrPPHighMult$
\end{tabular}
\end{table}

We compare our results for $(\D S/\D y) / (\D N_\mathrm{ch}/\D y)$ ratio based on experimental data, 
to the values obtained from the statistical hadron resonance gas model at the chemical freeze-out temperature of $T_\text{ch} = \unit[156]{MeV}$.
For the 0--10\% most central Pb--Pb collisions statistical model values are systematically
higher than our estimate, but in agreement at the 1--2$\sigma$ level. However the measured $(\D S/\D y) / (\D
N_\mathrm{ch}/\D y)$ values in minimum bias and high-multiplicity pp
collisions at $\sqrt{s} = \unit[7]{TeV}$ are  below the theory predictions for a
chemically equilibrated resonance gas at $T_{\text{ch}}=156\,\text{MeV}$, perhaps indicating 
that full chemical equilibrium is not reached in these
collisions. 
Here we note that,  interestingly, in pp collisions the estimated soft pion phase-space density exceeds unity. 
Finally, we have checked the dependence of our results on the relation between
one-dimensional and three-dimensional HBT radii, \Eq{eq:Rinv}, in \App{sec:RinvRoutRsideRlong}.
We found no significant change for Pb-Pb results, but pp entropy increased by $~10\%$,
which corresponds to  $1 \sigma$ deviation from the results above using \Eq{eq:Rinv}.

The precise knowledge of the total produced entropy in heavy ion
collisions and the entropy per final-state charged hadron is important for
constraining the bulk properties of the initial-state from the final state
observables~\cite{Kurkela:2018wud,Kurkela:2018xxd, Giacalone:2019ldn}. 
To determine  the
initial medium properties for high multiplicity pp and Pb--Pb
collisions, 
we performed simulations of  averaged initial conditions starting at $\tau_0 = 0.1\,\text{fm}/c$  with kinetic pre-equilibrium model
\kompost{}~\cite{Kurkela:2018wud,Kurkela:2018vqr,kompost_github} and viscous relativistic hydrodynamics code FluiduM~\cite{Floerchinger:2018pje}. Importantly, these calculations
take into account the produced entropy and  work done in
both the pre-equilibrium and hydrodynamic phases of the expansion~\cite{Bjorken:1982qr,Gyulassy:1983ub, Danielewicz:1984ww, Hwa:1985xg, Giacalone:2019ldn}.
We find that for simulations with the specific shear viscosity value $\eta/s=0.08$ the initial pre-equilibrium energy per unity rapidity is about three times larger than at the final state in  0--10\% most central Pb--Pb
collisions at $\sqrt{s_\mathrm{NN}} = \unit[2.76]{TeV}$, and approximately twice larger in high multiplicity pp collisions at $\sqrt{s} = \unit[7]{TeV}$.
At the time $\tau = \unit[1]{fm}/c$, the temperature in the center of the approximately equilibrated QGP fireball is about $T \approx \unit[400]{MeV}$ for Pb--Pb and $T \approx \unit[250]{MeV}$ for high-multiplicity pp collision systems.
Finally, we note that in our simulations of Pb--Pb collisions with $\eta/s=0.08$ only about 5\% of the total final entropy is
produced after $\tau = \unit[1]{fm}/c$, meaning that most of entropy production occurs in the pre-equilibrium phase.

\begin{acknowledgments}
The authors thank Stefan Floerchinger, Eduardo Grossi, and Jorrit Lion for
sharing the FluiduM package and useful discussions. A.M. thanks Giuliano Giacalone, Oscar Garcia-Montero, S\"oren Schlichting
and Derek Teaney for helpful discussions. Moreover, P.H. and K.R. thank Dariusz Miskowiec and Johanna Stachel for valuable discussions.
This work is part
of and supported by the DFG Collaborative Research Centre ``SFB 1225
(ISOQUANT)''.
\end{acknowledgments}

\appendix

\section{Relation between the 1D HBT radius $R_\mathrm{inv}$ and the 3D HBT radii $R_\mathrm{out}$, $R_\mathrm{side}$, $R_\mathrm{long}$}
\label{sec:RinvRoutRsideRlong}

A transformation of the three-dimensional Gaussian HBT radii from the
longitudinally co-moving system (LCMS) to the pair rest frame (PRF) only
affects the outwards direction according to $R_\mathrm{out}^\mathrm{PRF} =
\gamma R_\mathrm{out}^\mathrm{LCMS}$ where $\gamma = m_\mathrm{T}/m \equiv
\sqrt{m^2 + p_\mathrm{T}^2}/m$.
For a three-dimensional Gaussian parametrization of the source, the
one-dimensional distribution in 
radial distance from the origin
is not in general a
one-dimensional Gaussian. Therefore, there is no exact formula relating
the radius $R_\mathrm{inv}$ of the one-dimensional Gaussian parametrization of the source
and $R_\mathrm{out}$, $R_\mathrm{side}$, $R_\mathrm{long}$~\cite{Kisiel:2014upa}.

\begin{figure} 
\includegraphics[width=0.95\linewidth]{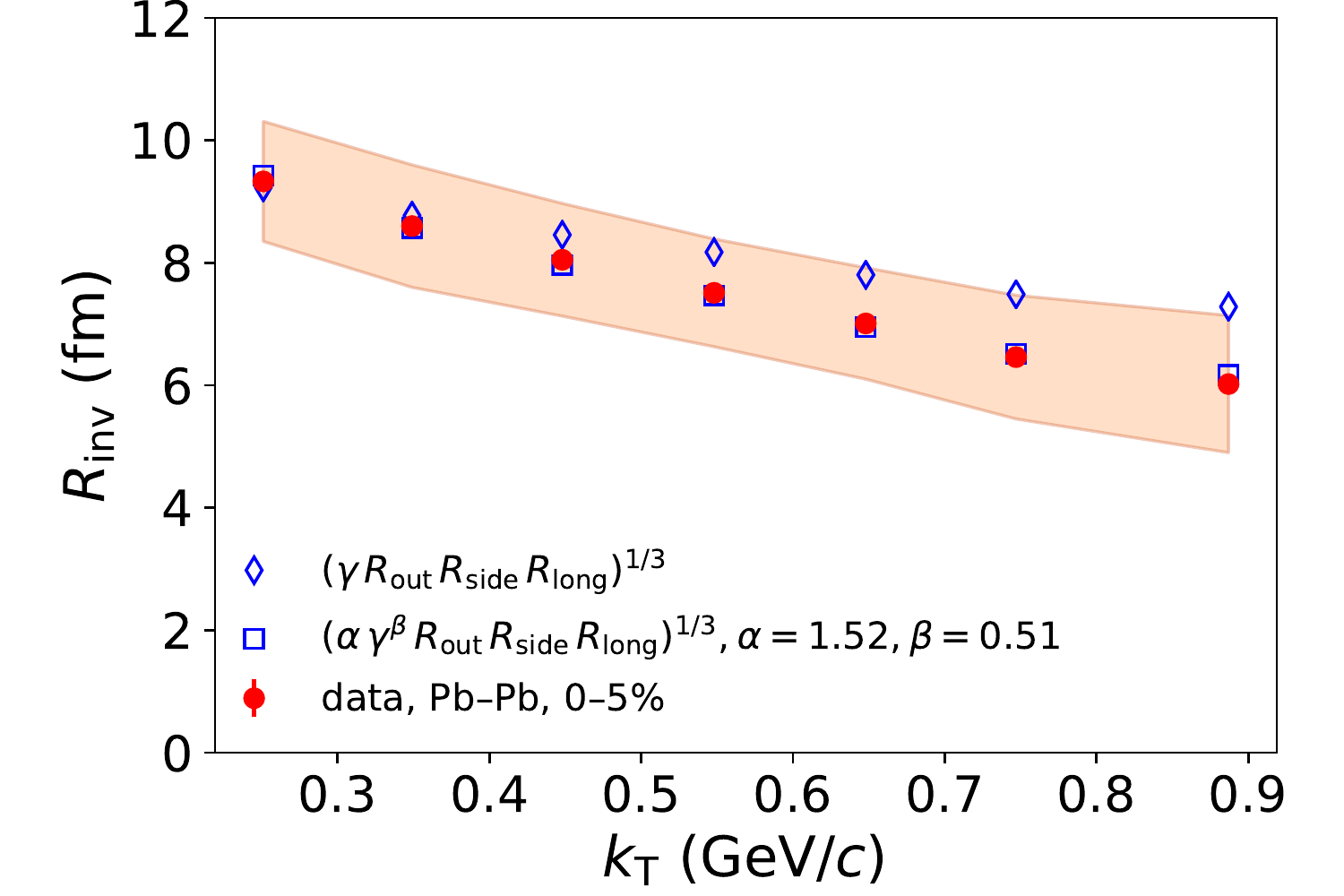}
\includegraphics[width=0.95\linewidth]{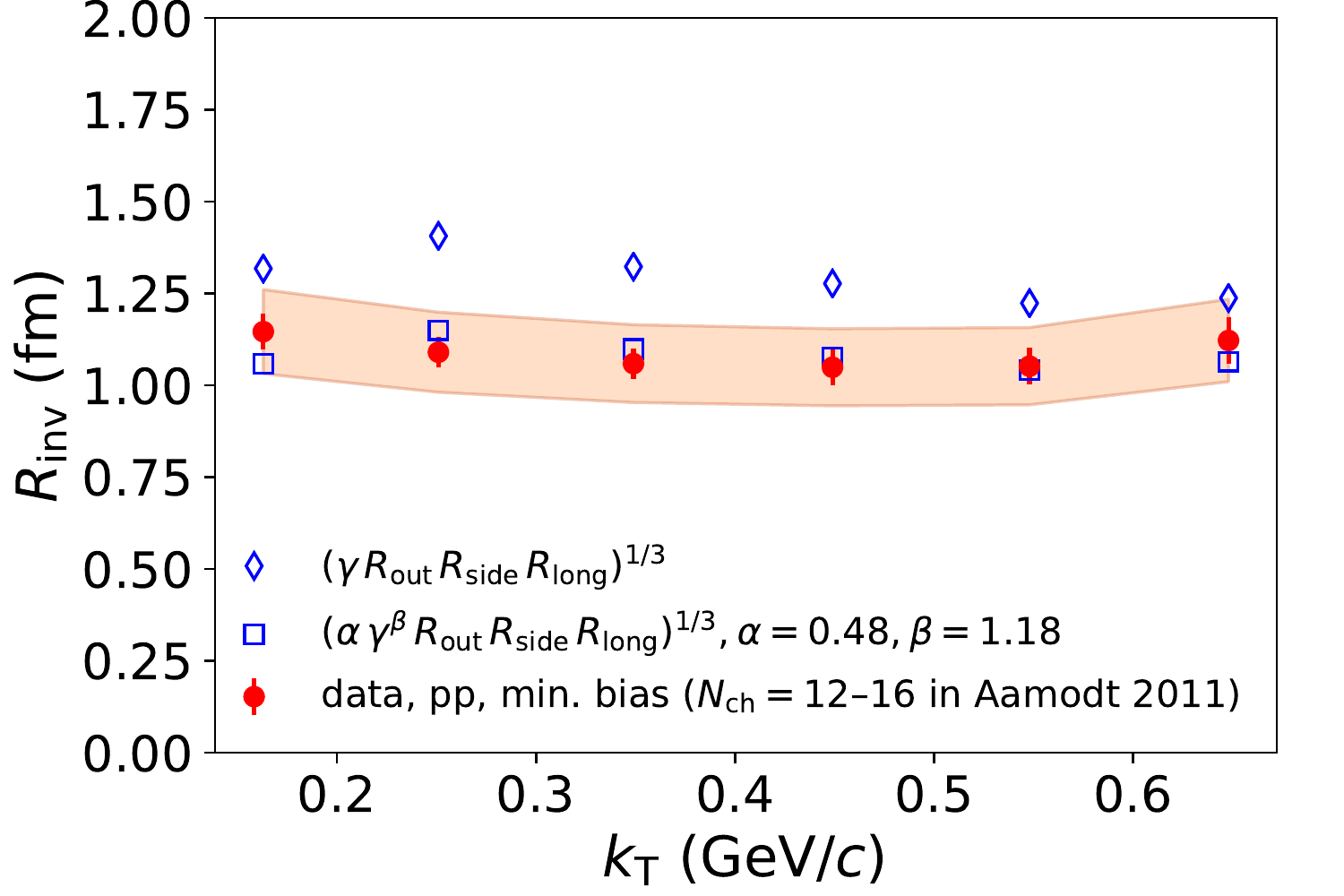}
\includegraphics[width=0.95\linewidth]{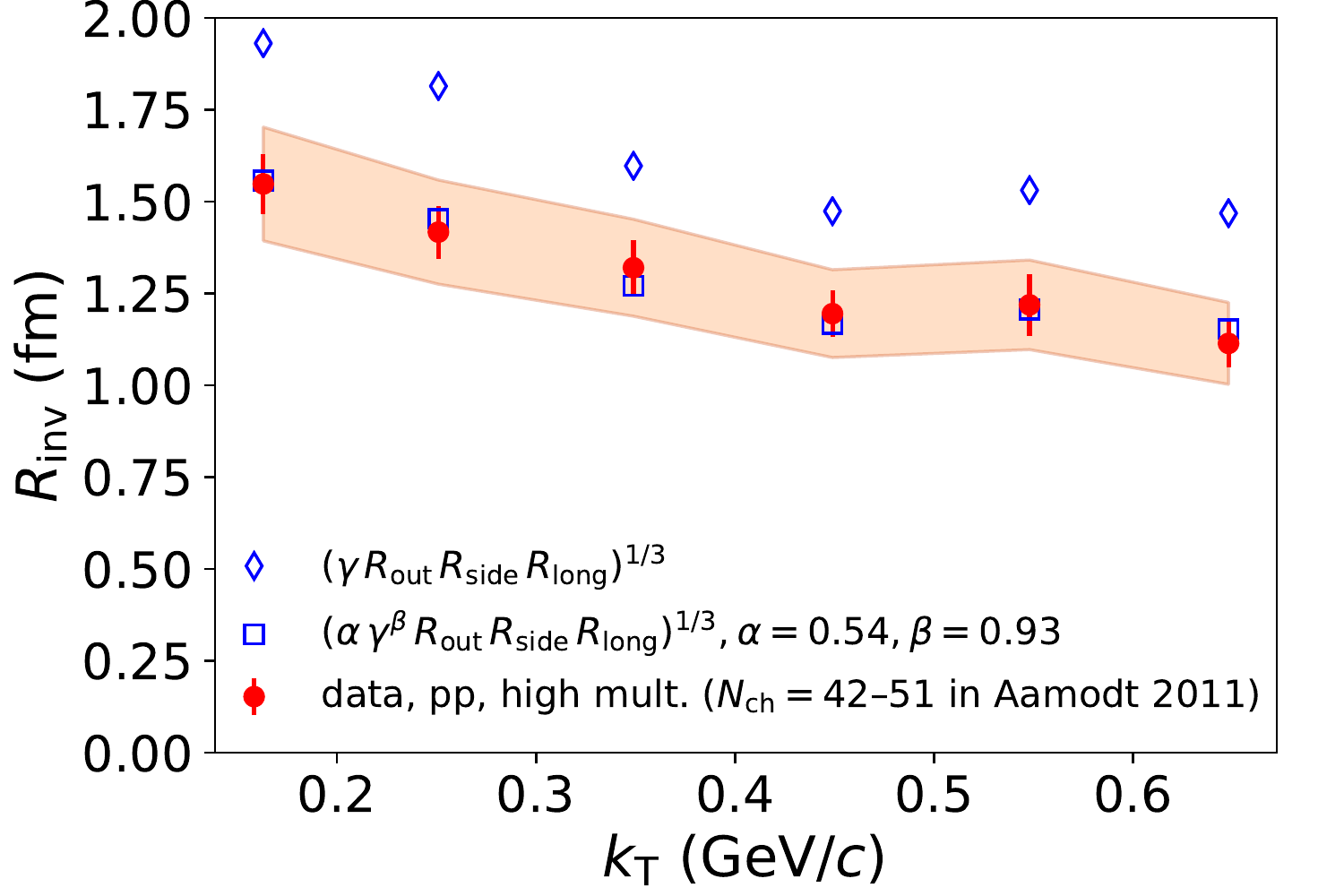}
\caption{\label{fig:Rinv_scaling} Comparison of measured one-dimensional HBT
radii $R_\mathrm{inv}$ as a function of the transverse pair momentum
$k_\mathrm{T}$ from the ALICE collaboration and $R_\mathrm{inv}$ values
calculated from measured three-dimensional HBT radii $R_\mathrm{out}$,
$R_\mathrm{side}$, $R_\mathrm{long}$ according to $R^3_\mathrm{inv} \approx
\alpha \gamma^\beta R_\mathrm{out} R_\mathrm{side} R_\mathrm{long}$. The 
bands reflect the uncertainties of the measured $R_\mathrm{inv}$ values.
The calculated $R_\mathrm{inv}$ values are shown for $\alpha = \beta = 1$ and for
the values of $\alpha$ and $\beta$ which fit the measured $R_\mathrm{inv}$
radii best. Results are shown for Pb--Pb collisions (centrality 0--5\%) at
$\sqrt{s_\mathrm{NN}} =
\unit[2.76]{TeV}$ \cite{Adam:2015vna} (upper panel), ``minimum bias'' pp collisions ($N_\mathrm{ch} = 12$--$16$ in (Aamodt 2011:~\cite{Aamodt:2011kd})) at $\sqrt{s} =
\unit[7]{TeV}$ \cite{Aamodt:2011kd} (middle panel), and high-multiplicity pp collisions ($N_\mathrm{ch} = 42$--$51$ in (Aamodt 2011:~\cite{Aamodt:2011kd})) at $\sqrt{s} =
\unit[7]{TeV}$ \cite{Aamodt:2011kd} (lower panel).} 
\end{figure}

The maximum phase space density $\mathcal{F}$ for a more general version of \Eq{eq:Rinv} of the form
\begin{equation}
R^3_\mathrm{inv} \approx 
h(\gamma) R_\mathrm{out} R_\mathrm{side} R_\mathrm{long}
\label{eq:RinvGeneral} 
\end{equation}
is given by
\begin{equation}
\mathcal{F} = \frac{h(\gamma)}{m_\mathrm{T}} \frac{(2 \pi)^{3/2}}{2J+1} 
\frac{1}{R_\mathrm{inv}^3} E\frac{\D^3 N}{\D^3 p}.
\label{eq:fmax_mod}
\end{equation}
In this section we use assume $h(\gamma) = \alpha \gamma^\beta$ and use data from ALICE \cite{Adam:2015vna,Aamodt:2011kd}
to determine the values of $\alpha$ and $\beta$ which best describe the relation
between the measured one-dimensional radii $R_\mathrm{inv}$ and the three-dimensional
radii $R_\mathrm{out}$, $R_\mathrm{side}$, $R_\mathrm{long}$.
The results are shown in Fig.~\ref{fig:Rinv_scaling}. 
The best fit values for $\alpha$ and $\beta$ turn out be significantly different between Pb--Pb and pp collisions. We then use these values to calculate maximum phase space densities according to
\Eq{eq:fmax_mod}. The corresponding results for the entropy and the
entropy per final-state charged particle are:

\begin{table}[h!]
\begin{tabular}{ccc}
system & $\D S/\D y$ & $(\D S/\D y) / (\D N_\mathrm{ch}/\D y)$ \\
\hline 
Pb--Pb, 0--10\%& $\dSdyGen \pm \dSdyGenErr$ & $\SoverNchGen \pm \SoverNchGenErr$ \\
pp minimum bias &  $\dSdyGenPPMinBias \pm \dSdyGenPPMinBiasErr$ & $\SoverNchGenPPMinBias \pm \SoverNchGenErrPPMinBias$ \\
pp high mult. & $\dSdyGenPPHighMult \pm \dSdyGenPPHighMultErr$ & $\SoverNchGenPPHighMult \pm \SoverNchGenErrPPHighMult$
\end{tabular}
\end{table}
We note that for Pb--Pb collisions the entropy estimates increases only very slightly. For pp collisions, the values for the entropy are higher than our standard results obtain using $\alpha = \beta = 1$, but they agree at the $1\sigma$ level.

\section{Two-component Glauber model \label{sec:Glauber}}
In this section we recap the details of the two-component Glauber model used to generate initial conditions for  \kompost{} evolution.
The nuclear charge density distribution of lead nuclei is parametrized by Wood-Saxon distribution~\cite{Miller:2007ri}
\begin{equation}
  \rho(\vec{r}) = \rho_0 \frac{1}{1+ \exp\left( \frac{|\vec{r}|-R}{a} \right) }
,\end{equation}
where for our purposes we will choose $\rho_0$ such that the total volume integral of $\rho$ is equal to the number of nucleons $N_A=208$. Then
$\rho_0= \unit[0.160391]{fm}^{-3}$, $R=\unit[6.62]{fm}$, and $a=\unit[0.546]{fm}$.
For Lorentz contracted nuclei, the longitudinal direction can be integrated out to obtain the density per unit transverse area
\begin{equation}
  T(\vec r_\perp) = \int_{-\infty}^\infty \rho\left( \vec{r}_\perp, z \right) dz 
.\end{equation}
Then the collision probability of two nuclei with $N_A$ and $N_B$ nucleons is given by
\begin{equation}
  \frac{\D N^\text{coll}(\vec r, \vec b)}{\D^2 \vec r} =  T_A(\vec r) T_B(\vec r-\vec b) \sigma^{NN}_{\text{inel}}
,\end{equation}
where the radius is implicitly assumed to be in the transverse plane and $\sigma^{NN}_{\text{inel}}=\unit[6.4]{fm}^2$ is the inelastic nucleon-nucleon cross-section. The number of participant nucleons per transverse area is given by
\begin{align}
  \frac{\D N^\text{part}(\vec r, \vec b)}{\D^2 \vec r} &=  T_A(\vec r) \left[1-\left(1-T_B(\vec r-\vec b) \sigma^{NN}_{\text{inel}}/N_B \right)^{N_B}\right]\nonumber\\
                                         &+  T_B(\vec r-\vec b) \left[1-\left(1-T_A(\vec r) \sigma^{NN}_{\text{inel}} /N_A \right)^{N_A}\right]
.\end{align}
These probabilities are combined in the two-component Glauber model~\cite{Miller:2007ri,Abelev:2013qoq} 
where $\alpha$ is an adjustable parameter
\begin{equation}
(s\tau)_0 =\kappa_s \left(\frac{1-\alpha}{2} \frac{\D N^\text{part}(\vec r, \vec b)}{\D^2r}
+ \alpha \frac{\D N^\text{coll}(\vec r, \vec b)}{\D^2r}\right)\label{eq:2comp}
\end{equation}
We use $\alpha=0.128$, which is the same value as  in ALICE publication~\cite{Abelev:2013qoq}, but with
different parametrization of \Eq{eq:2comp}, namely
$\alpha = \frac{1-f}{1+f}$.

\bibliography{entropy_lhc}

\end{document}